\documentclass[twocolumn]{aastex631}
\usepackage{amssymb}
\usepackage{amsmath}
\usepackage{comment,ulem}
\usepackage{graphicx}
\usepackage{tabularx}
\usepackage{tikz-feynman}

\usepackage{lipsum}

\shorttitle{The effects of self-interacting bosonic dark matter on neutron star properties}
\shortauthors{Giangrandi et al.}

\graphicspath{{./}{figures/}}

\begin{document}

\title{The effects of self-interacting bosonic dark matter on neutron star properties}

\author[0000-0001-9545-466X]{Edoardo Giangrandi}
\affiliation{CFisUC, Department of Physics, University of Coimbra, Rua Larga P-3004-516, Coimbra, Portugal}

\author[0000-0001-5854-1617]{Violetta Sagun}
\affiliation{CFisUC, Department of Physics, University of Coimbra, Rua Larga P-3004-516, Coimbra, Portugal}

\author[0000-0002-4947-8721]{Oleksii Ivanytskyi}
\affiliation{Incubator of Scientific Excellence---Centre for Simulations of Superdense
Fluids, University of Wrocław, 50-204, Wroclaw, Poland}

\author[0000-0001-6464-8023]{Constança Providência}
\affiliation{CFisUC, Department of Physics, University of Coimbra, Rua Larga P-3004-516, Coimbra, Portugal}

\author[0000-0003-2374-307X]{Tim Dietrich}
\affiliation{Institut für Physik und Astronomie, Universität Potsdam, Haus 28,
Karl-Liebknecht-Str. 24/25, Potsdam, Germany}
\affiliation{Max Planck Institute for Gravitational Physics (Albert Einstein Institute), Am Mühlenberg 1, Potsdam D-14476, Germany}

\begin{abstract}
We propose a model of asymmetric bosonic dark matter (DM) with self-repulsion. By adopting the two-fluid formalism, we study different DM distribution regimes, either, fully condensed inside the core of a star, or, otherwise, distributed in a dilute halo around a neutron star (NS). We show that for a given total gravitational mass, DM condensed in a core leads to a smaller radius and tidal deformability compared to a pure baryonic star. This effect may be interpreted as an effective softening of the equation of state. On the other hand, the presence of a DM halo increases the tidal deformability and total gravitational mass. As a result, an accumulated DM inside compact stars could mimic an apparent softening/stiffening of strongly interacting matter EoS and constraints we impose on it at high densities.

We limit the model parameter space by confronting the cross-section of the DM self-interaction to the constraint extracted from the analysis of the Bullet Cluster. Furthermore, from the performed analysis of the effect of DM particles, interaction strength, and relative DM fractions inside NSs we obtained a rigorous constraint on model parameters. To identify its impact on NSs we consider the DM fraction may reach up to 5\%,  which could be considered too high in several scenarios. Finally, we discuss several pieces of smoking gun evidence of the presence of DM that is free from the abovementioned degeneracy between the effect of DM and properties of strongly interacting matter. These signals could be probed with future and ongoing astrophysical and gravitational wave (GW) surveys.
 
\end{abstract}

\keywords{Neutron Stars(1108) --- Dark Matter(353) --- Gravitational Waves(678)}

\section{Introduction} 
\label{sec:intro}

Since the first detection of the binary neutron star (NS) merger, GW170817 \citep{LIGOScientific:2017vwq}, which was accompanied by the observation of electromagnetic signals originating from the same source, GRB170817A and AT2017gfo  \citep{LIGOScientific:2017ync}, we have been witnessing exciting breakthroughs in our understanding of compact stars and their merger dynamics. In fact, gravitational wave (GW) astronomy and multi-messenger astrophysics became new tools to extract information about the internal structure of NSs from GW and electromagnetic observations \citep{Bauswein:2017vtn, Annala:2017llu,Hinderer:2018pei}. Thus, from the combination of the analyses of the GW170817 signal measured by the advanced LIGO and advanced Virgo detectors, the constraint on the tidal deformability parameter of NS matter  $\Lambda_{1.4} \leq 800$ was extracted \citep{Abbott_2018}. The second binary NS merger event, GW190425 \citep{LIGOScientific:2020aai}, provided constraints consistent with GW170817, but due to its lower signal-to-noise ratio did not deepen our knowledge about the NS Equation of State (EoS). In addition to GW observations, also X-ray observations by NICER \citep{Miller_2019,Miller:2021qha,Riley:2019yda,Riley:2021pdl,Raaijmakers:2019dks}, radio measurements of the heaviest pulsars, e.g. PSR J0348+0432 of mass $2.01\pm 0.04$ $M_{\odot}$ \citep{PSRj03480432Article}, PSR J0740+6620 of $2.08^{+0.07}_{-0.07}$ $M_{\odot}$ \citep{Fonseca:2021wxt} and optical observations of the \textit{black widow} pulsars, e.g., PSR J1810+1744 of $2.13\pm 0.04$ $M_{\odot}$ \citep{Romani:2021xmb}, and PSR J0952-0607 of $2.35\pm 0.17$ $M_{\odot}$ \citep{Romani:2022jhd}
constrain properties of NSs.

While all the mentioned analyses and models assume that NSs are embedded in a pure vacuum and do not contain dark matter (DM), they, indeed, could accumulate a sizable amount of DM in their interior and surroundings. Due to high compactness, NSs can effectively trap DM particles, which will rapidly thermalize and become accrued inside the stars, altering their properties. The presence of DM affects the internal structure and compactness of compact stars. Thus, as it was shown, e.g., \citet{Ciarcelluti:2010ji,2018PhRvD..97l3007E,2019JCAP...07..012N,PhysRevD.102.063028,2020MNRAS.495.4893D} and~\citet{Sagun:20224z}, DM may either form an extended halo or a dense core inside a NS. Depending on the mass of DM particles, their self-interaction strength, and its relative abundance inside the star, one of the abovementioned scenarios takes place. 
Since DM halos are invisible for typical astrophysical observations, we would see only the baryonic matter (BM) radius, independent of the fact that the outermost radius can extend further than its BM component~\citep{Rafiei_Karkevandi_2022}. On the contrary, a DM core formation will lead to a reduction in the NS radius. 

Moreover, DM will affect tidal deformability parameters and the merger dynamics \citep{Ellis:2017jgp,Bezares:2019jcb,Bauswein:2020kor,Leung:2022wcf}. Nowadays, while there are studies investigating possible alternative scenarios beyond \textit{standard} compact binary mergers described by general relativity in pure vacuum \citep{LIGOScientific:2018dkp}, the models used to analyze GW signal do not account directly for DM. 

Thus, to understand the effect of DM on the coalescence of NSs, numerical-relativity simulations for different DM fractions, particle mass, and interaction strength are required.
As a step in this direction, there have been the first two-fluid 3D simulations of coalescing binary NS admixed with DM with the following studies of GW emission of the merger remnant, e.g.~\citet{Bauswein:2020kor} and~\citet{Emma:2022xjs}.
By considering different binary masses and EoSs,~\citet{Bauswein:2020kor} showed that the GW frequency of the orbiting DM components scales with the compactness of NSs. Moreover, the relations between the DM GW frequency and the dominant post-merger GW frequency of the stellar fluid or the tidal deformability were found, which opens a possibility to probe the EoS effects during the binary inspiral. \citet{Emma:2022xjs} studied the effect of mirror DM concentrated inside the core on the deceleration of the inspiral phase, as well as on a modification of the ejecta and debris disk formation. 

Depending on whether DM has particle-antiparticle asymmetry, we will refer to it as asymmetric or symmetric matter. Symmetric DM particles can self-annihilate leaving a possibility of its detection via X-ray, $\gamma$-ray, or neutrino telescopes \citep{KouvarisThermalEvolution}. Moreover, as studied in \citet{2012PhLB..711....6P}, self-annihilating DM in the inner regions of NSs may have a significant impact on the kinematic properties, namely, velocity kicks and rotation patterns.

Another possible effect of DM particle annihilation inside the NS core is related to the late-time heating, which could be detected from observations of the surface temperature of the old part of the NS population \citep{2010PhRvD..81l3521D,Hamaguchi:2019oev}. Unfortunately, nowadays, our database of old NSs is still quite limited. 

Contrary to the annihilating DM, asymmetric DM will become accumulated inside a star. Models that consider such scenario should allow old NSs to exist. Especially, it is important for bosonic DM particles, which at zero temperature could form Bose-Einstein condensate (BEC) leading to the gravitational collapse of the bosonic DM to a black hole \citep{Kouvaris}.

Light DM particles, such as axions, could contribute as an additional cooling channel in compact stars. Thus, in the NS core axions can be produced either in nucleon bremsstrahlung or in Cooper pair breaking and formation processes \citep{PhysRevD.93.065044,Sedrakian:2018kdm,Buschmann:2021juv}, causing an alteration of the surface temperature and thermal evolution of a star.  In addition, most of the existing models are constrained by the results of neutrino emission coming from the supernova observation SN 1987A \citep{Chang:2018rso} and existing NS cooling data. The results of NS merger simulations  \citep{Dietrich:2019shr} show that axions produced in nucleon-nucleon bremsstrahlung do not lead to a measurable change in the emitted GW signal, ejecta mass, as well as the temperature profile of the merger remnant.

The fraction of DM in the compact star interior depends on different factors. Thus, DM can be captured by NSs from the surrounding medium. Following scattering processes, the kinetic energy of DM particles is transferred to the star \citep{Kouvaris:2013awa,Bell:2020jou,Bell:2020obw}, and becomes gravitationally bound with a star. 
The amount of DM accrued by an ordinary accretion throughout a stellar evolution will depend on the position of the considered NS in the Galaxy \citep{2010PhRvD..82f3531K}. As the DM density in the Galactic Center is many orders of magnitude greater than in its arms, we may expect a higher DM fraction in compact stars toward the Milky Way center \citep{2020Univ....6..222D}. Furthermore, NSs in globular clusters may contain a significant amount of DM \citep{Bertone:2007ae}. Moreover, we should not forget that NS is the final stage of star's evolution preceded by the progenitor, main-sequence star, and supernova explosion with the formation of a proto-NS. These and all other mechanisms are discussed in detail in \citet{Rafiei_Karkevandi_2022}.
As was estimated by \citet{PhysRevD.102.063028}, the amount of accumulated DM in the most central part of the Galaxy accrued by a spherically symmetric accretion scenario during the main-sequence star and equilibrated NS stages is 0.01\%. However, additional scenarios could lead to high DM factions inside compact stars, e.g., DM production during a supernova explosion, accretion of DM clumps formed at the early stage of the universe, or initial star formation on a preexisting dark core. As we are interested in identifying possible signatures on NSs, we have concentrated our study on fractions up to 5\%, which cannot be accumulated by a spherically symmetric DM accretion followed by thermalization via the interaction with BM, and requires the alternative mechanisms mentioned above.

Moreover, some local non-homogeneity of DM distribution may contribute to an increase in DM fraction, leading even to dark compact objects \citep{Dengler:2021qcq} and dark stars \citep{2015PhRvD..92f3526K,Maselli:2017vfi}. 

Since DM properties are still unknown, different models have been employed, considering its fermionic  \citep{Goldman:2013qla,Gresham:2017zqi,PhysRevD.102.063028} and bosonic \citep{Colpi:1986ye,Petraki:2013wwa,Rafiei_Karkevandi_2022} nature. As it was discussed by \citet{Bramante:2013hn} to be consistent with the observations of old NSs, bosonic DM has to be either self-interacting, decaying, or self-annihilating. Considering asymmetric DM a repulsive self-interaction is required due to zero degeneracy pressure. At the moment when accumulated bosonic asymmetric DM exceeds the Chandrasekhar mass, nothing can prevent its gravitational collapse and the formation of a black hole inside the NS, which could potentially disrupt the star \citep{Kouvaris, Zurek:2013wia}. 

Using an analog of visible matter and the Standard Model particles, we see that all interactions have an exchange character, an interaction between particles occurs due to an exchange of a mediator, e.g., the interaction between nucleons is mediated by pions. In the present article, we extend this approach for a dark sector by formulating a model of self-interacting asymmetric bosonic DM, which includes vector interaction mediated by a real $\omega$-field coupled to the scalar one. We model DM-admixed compact stars by considering the mixed system of two fluids with different relative fractions. 

The assumption of cold self-interacting DM provides a good agreement with the large-scale structure of the universe and cosmology. It reconciles the success of the cold DM (CDM) model with the non-observation of cuspy density profiles of dwarf galaxies predicted by CDM \textit{N}-body simulations and known as the core-cusp problem \citep{Moore:1994yx}. In comparison to an alternative mechanism to flatten the central density profile by supernova-driven episodes of gas removal,  self-interacting DM is a more favorable one \citep{Burger:2022jid}. On the other hand, from the observed mass profiles of galaxies \citep{Ahn:2004xt} and Bullet Cluster observations \citep{Clowe_2006, Randall:2008ppe} the DM self-interaction cross section per unit mass has an upper limit of $\sigma / m <$1.25 cm$^{2}$ g$^{-1}$ (68\% confidence level). The DM model considered in this paper is in line with the above assumption, and therefore, provides consistency with the state of the art of modern cosmology. Moreover, below we explicitly account for the above constraint on $\sigma/m$ in order to limit the model parameters.
An implication of the proposed EoS and tests against astrophysical and GW observations are performed in this work. 

The paper is organized as follows. In Section \ref{sec:EoS}, we present models for the BM and DM components, with a detailed derivation provided in the Appendix \ref{appA:FullLagrangian}. Section \ref{sec:MIXEoS} is dedicated to the equilibrium configurations of DM-admixed compact stars. In Section \ref{sec:TID}, we discuss how the speed of sound and the tidal deformability are affected by the presence of DM. In Section \ref{sec:Results} the main results are presented, including the constraints on mass and interaction scale of DM particles. In Section \ref{sec:Discussions}, we discuss the smoking gun signals of the presence of DM that could be tested in the nearest future before concluding in Section \ref{sec:Concl}. In Appendix \ref{appA:FullLagrangian}, we show the full derivation of the DM EoS, with a focus on the effective speed of sound for a DM-admixed NS in Appendix \ref{appB:SpeedOfSound}. In Appendix \ref{appC:Scan}, we show the scan over the model parameters and the obtained constraints.
Throughout the article, we utilize a unit system in which $\hbar=c=G=1$.

\section{Models of dark and baryonic matter} 
\label{sec:EoS}

\subsection{DM EoS}
\label{subsec:DMEoS}
We consider the model of massive spinless DM particles carrying a conserved charge. Such particles are described by a complex scalar field, have mass $m_\chi$ and chemical potential $\mu_\chi$. At sufficiently low temperatures bosonic DM exists in the form of the BEC. In the absence of interaction such BEC has zero pressure and is mechanically unstable against gravitational compression. We stabilize the BEC of DM by introducing repulsive interaction mediated by a real vector field coupled to the scalar one. The minimal Lagrangian representing this model is given in the Appendix \ref{appA:FullLagrangian}. It is equivalent to a massive\textit{ U}(1) gauge theory of scalar particles, i.e., scalar electrodynamics with massive photons. This Lagrangian implies a Noether current corresponding to the invariance of action with respect to global \textit{U}(1) transformations. If the vector field was not a Yukawa but a gauge one, local \textit{U}(1) symmetry would also be respected and another Noether current could be introduced \citep{Brading:2000hc}. Given a quantum treatment, expectation values of these two currents produce the same conserved charge, which is not the case within the used mean-field approximation corresponding to a classical treatment of the vector field (see Appendix \ref{appA:FullLagrangian} for details). We use the Noether current resulting from global \textit{U}(1) transformations, which leave the action invariant even at the mean-field level.

In this work, we assume the vanishing temperature of the DM, being totally converted to the BEC. In the considered case thermal fluctuations are suppressed and mean-field approximation can be applied in order to derive the corresponding EoS. Chemical potentials of the BM and DM components of NS scale proportionally (for more details see Section \ref{sec:MIXEoS}).  This significantly simplifies solving two coupled Tolman–Oppen-
heimer–Volkoff (TOV)–like equations for BM and DM components, as shown by \citet{PhysRevD.102.063028}. Therefore, it is convenient to formulate the DM EoS in the grand canonical ensemble (GCE), where $\mu_\chi$ is an independent variable. Appendix \ref{appA:FullLagrangian} includes details of the corresponding derivation for the interval of physical values of $\mu_\chi\in[0,\sqrt{2}m_\chi]$ performed in the locally flat spacetime, provided by small gradients of metrics and absence of the anisotropy issues (see \citet{Rafiei_Karkevandi_2022} for details). The corresponding pressure and energy density are
\begin{eqnarray}
\label{EqI}
p_\chi&=&\frac{m_I^2}{4}
\left(m_\chi^2-\mu_\chi\sqrt{2m_\chi^2-\mu_\chi^2}\right),\\
\label{EqII}
\varepsilon_\chi&=&\frac{m_I^2}{4}
\left(\frac{\mu_\chi^3}{\sqrt{2m_\chi^2-\mu_\chi^2}}-m_\chi^2\right),
\end{eqnarray}
for $\mu_\chi\in[m_\chi,\sqrt{2}m_\chi]$ and $p_\chi=\varepsilon_\chi=0$ for $\mu_\chi\in[0,m_\chi]$. The parameter $m_I$ has the unit of mass and controls the interaction strength. It is proportional to the vector meson mass $m_\omega$ and inversely proportional to its coupling $g$. Thus, large $m_I$ corresponds to weak interaction and vice versa. At a first glance, the present EoS in the weak coupling regime paradoxically leads to an infinite pressure due to $m_I\rightarrow\infty$. This, however, is not the case since in the considered regime chemical potential of the DM BEC $\mu_\chi$ coincides with its mass $m_\chi$ leading to the vanishing of the brackets in Eqs.~\eqref{EqI} and~\eqref{EqII}. In the case of $p_\chi$ the bracket vanishes faster than $m_I^2$ yielding to a zero pressure $\sim m_I^{-2}$, while for $\varepsilon_\chi$ the bracket behaves as $\sim m_I^{-2}$ providing a finite energy density of the DM BEC $m_\chi n_\chi$. In the strong coupling regime, $m_I\rightarrow0$ chemical potential of DM converges to $\sqrt{2}m_\chi$. As a result, the bracket in Eq.~\eqref{EqI} becomes equal to $m_\chi^2$ and the pressure vanishes as $m_I^2m_\chi^2/4$. The corresponding bracket in Eq.~\eqref{EqII} diverges as $\sim m_I^{-2}$ leading to finite energy density $\sqrt{2}m_\chi n_\chi$. Remarkably, weak and strong coupling limits of the present EoS are similar, since DM pressure vanishes in both these cases. At $m_I\rightarrow\infty$ it is due to the absence of repulsion. The limit $m_I\rightarrow0$ is equivalent to the case of the massless vector field, which does not have a nontrivial mean-field solution needed to stiffen the EoS. Detailed analysis of the weak and strong coupling limits of the present EoS is performed in Appendix \ref{appA:FullLagrangian}.

\begin{figure}
\includegraphics[width=1.15\columnwidth]{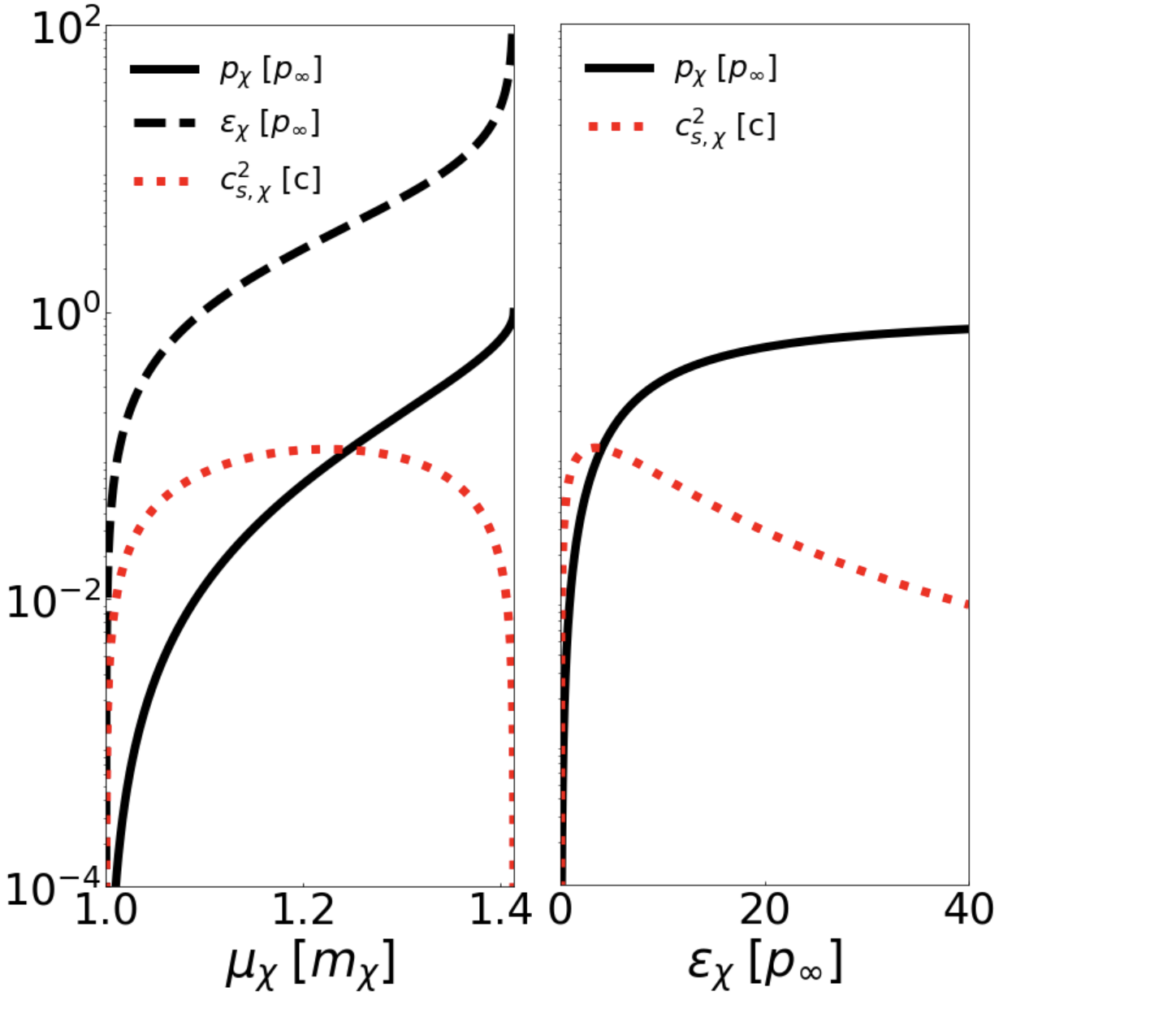}
\caption{{\bf Left panel:} Scaled pressure $p_\chi/p_\infty$ (black solid curve), energy density $\varepsilon_\chi/p_\infty$ (black dashed curve) and speed of sound squared $c_{s,\chi}^2$ (red dotted curve) of DM as functions of its chemical potential $\mu_\chi$ given in units of $m_\chi$. {\bf Right panel:}
Scaled pressure $p_\chi/p_\infty$ (black solid curve) and speed of sound $c_{s,\chi}^2$ (red dotted curve) of DM as functions of scaled energy density $\varepsilon_\chi$ given in units of $p_\infty$.
}
\label{fig1}
\end{figure}

A remarkable feature of the present EoS is that at infinite density its pressure is limited by the value $p_\infty=m_I^2m_\chi^2/4$. This regime is reached at $\mu_\chi=\sqrt{2} m_\chi$. Thus, the compressibility of DM vanishes at asymptotically high densities regardless of $m_\chi$ and $m_I$. The same conclusion holds for the speed of sound $c_{s,\chi}^2=dp_\chi/d\varepsilon_\chi$. In other words, high-density configurations of bosonic DM are gravitationally unstable at any strength of the repulsive interaction. The left panel of Fig.~\ref{fig1} shows the pressure, energy density, and speed of sound of the considered DM EoS as functions of the corresponding chemical potential. It is worth mentioning, that the square of the speed of sound is limited from above by the value $1/9$, which is reached at $\mu_\chi=\sqrt{3/2}~m_\chi$ and does not depend on $m_\chi$ and $m_I$. Thus, $c_{s,\chi}^2$ is bounded by quite small values and corresponds to the soft EoS of DM. The right panel of Fig.~\ref{fig1} shows this EoS as a function of energy density.

Observational data on the colliding clusters of galaxies 1E 0657-56 (the Bullet Cluster) enable probing dynamics of the DM fluid on the cosmological scale. This dynamic is determined by the DM self-interaction, which is controlled by the corresponding cross section $\sigma_\chi$. Evaluation of this quantity requires the invariant matrix element of scattering of two on-shell DM particles from the initial state with three momenta ${\bf k}_1$ and ${\bf k}_2$ to the final one with momenta ${\bf k}_1'$ and ${\bf k}_2'$. At the tree level, this matrix element includes contributions from the $t$- and $u$-channels
\begin{eqnarray}
\label{matrix_element}
i\mathcal{M}&=&
\begin{array}{c}
\begin{tikzpicture}
\begin{feynman}
\vertex (a);
\vertex [left = of a]  (b) {${\bf k}_1$};
\vertex [right = of a] (c) {${\bf k}_1'$};
\vertex [below = of a] (d);
\vertex [left = of d]  (e) {${\bf k}_2$};
\vertex [right = of d] (f) {${\bf k}_2'$};
\diagram{(b) -- [fermion] (a) -- [fermion] (c)};
\diagram{(a) -- [boson, edge label = $q$] (d)};
\diagram{(e) -- [fermion] (d) -- [fermion] (f)};
\end{feynman}
\end{tikzpicture}
\end{array}\nonumber\\
&+&
\begin{array}{c}
\begin{tikzpicture}
\begin{feynman}
\vertex (a);
\vertex [left = of a]  (b) {${\bf k}_1$};
\vertex [below right = of a] (c) {${\bf k}_2'$};
\vertex [below = of a] (d);
\vertex [left = of d]  (e) {${\bf k}_2$};
\vertex [above right = of d] (f) {${\bf k}_1'$};
\diagram{(b) -- [fermion] (a) -- [fermion] (c)};
\diagram{(a) -- [boson, edge label = $q$] (d)};
\diagram{(e) -- [fermion] (d) -- [fermion] (f)};
\end{feynman}
\end{tikzpicture}
\end{array}\,.
\end{eqnarray}
These diagrams are formalized by the Feynman rules of scalar quantum electrodynamics with massive photons. Each vertex corresponds to a factor $-ig$. The annihilation of a particle with four-momentum $k$ and the creation of another one with four-momentum $k'$ in a vertex produces the factor $-k_\mu-k_\mu'$, while in the case of antiparticles, the momenta entering this factor have the opposite signs. Wavy lines stand for the vector field propagator $-ig^{\mu\nu}(q^2-m_\omega^2)^{-1}$ with $q$ being a transferred momentum. The conservation of energy and momentum ensures that in the \textit{t}- and \textit{u}-channels, which are represented by the upper and lower graphs in Eq. (\ref{matrix_element}), squared transferred momentum coincides with the Mandelstam variables $t=(k_1-k_1')^2=(k_2-k_2')^2$ and $u=(k_1-k_2')^2=(k_2-k_1')^2$, respectively.
The Lorentz indexes $\mu$ and $\nu$ appearing in the vertexes and in the vector field propagator are dummies. With this, we arrive at
\begin{eqnarray}
\mathcal{M}=g^2\left[\frac{(k_1+k_1')(k_2+k_2')}{t-m_\omega^2}+
\frac{(k_1+k_2')(k_2+k_1')}{u-m_\omega^2}\right].\quad
\end{eqnarray}
In the BEC case, three momenta of the incoming and outgoing particles vanish, which produces $4m_\chi^2$ in the numerator of each fraction in the previous expression and corresponds to $t=u=0$. This yields $\mathcal{M}=-8 m_\chi^2/m_{I}^2$ and differential cross section of the DM self-interaction $d\sigma_\chi/d\Omega=|\mathcal{M}|^2/64\pi^2E_{\rm cm}$, where $E_{\rm cm}=2m_\chi$ is the center-of-mass energy of the incoming particles with ${\bf k}_1={\bf k}_2=0$. This differential cross section is independent of the angle variables. Therefore, the total one is obtained by multiplying $d\sigma_\chi/d\Omega$ by $4\pi$. Finally, we obtain
\begin{eqnarray}
\frac{\sigma_\chi}{m_\chi}=\frac{2 m_\chi}{\pi m_I^4}.
\end{eqnarray}

Numerical simulations of the Bullet Cluster combined with the results from X-ray, strong and weak lensing, optical observation set an upper limit on this ratio \cite{Randall:2008ppe}. Within the 68 \% confidence interval $\sigma_\chi/m_\chi<$1.25 cm$^2$ g$^{-1}$,
while assuming equal mass-to-light ratios in the subcluster and the main cluster prior to the merger yields even more stringent constraint 
$\sigma_\chi/m_\chi<$0.7  cm$^2$ g$^{-1}$.
In order to keep the parameter space of the model as wide as possible, we use a more relaxed version of this constraint.
Thus, we require
\begin{eqnarray}
\label{constraint}
m_I~[{\rm MeV}]>18.24~\sqrt[4]{m_\chi~[{\rm MeV}]}.
\end{eqnarray}
This requirement obviously discredits the strong coupling limit $m_I\rightarrow0$ with respect to the above cosmological constraint on the DM self-interaction cross section. Further analysis is limited to the region of the model parameter space, which respects Eq. (\ref{constraint}).

\subsection{Baryon matter EoS}
\label{subsec:BMEoS}

In order to thoroughly study the impact of DM on compact stars made of mostly BM, we consider two EoSs of different stiffness. One of them is the induced surface tension (IST) EoS, formulated on the basis of the hard-core approach. Thus, nucleons are characterized by an effective hard-core radius that provides a short-range repulsion between the particles of different species. This part of the model was fixed from the fit of heavy-ion collision data~\citep{Sagun:2017eye}, while the IST contribution was implemented by accounting for an interparticle interaction at high density. The corresponding parameters were fitted to reproduce the nuclear matter ground-state properties, correct behavior of the nuclear liquid-gas phase transition and its critical point~\citep{Sagun:2016nlv} and proton flow constraint~\citep{Ivanytskyi:2017pkt}. Furthermore, in~\citet{Sagun2019IST} the model was generalized to describe NSs showing a big application range of the unified IST approach~\citet{Sagun:2018sps}. In the present work, we consider the Set B described in detail in~\citet{NSOscillationsEoS}, while the crust is modeled in a simplified way by the polytropic EoS with adiabatic index $\gamma=4/3$. This model parameterization reproduces GW170817 and GW190425  tidal deformability limit~\citep{Abbott_2018, LIGOScientific:2020aai}, NICER mass-radius measurements~\citep{Miller_2019,Miller:2021qha,Riley:2019yda,Riley:2021pdl,Raaijmakers:2019dks}, as well as the maximum mass constraint.

In addition, we consider the DD2 EoS \citep{Typel1999,Typel2009} with and without $\Lambda$ hyperons. The DD2 is a mean-field relativistic nuclear model with density-dependent couplings, whose parameters were fitted to the ground-state properties of nuclei. Hyperons have been included in several works. In the present study, the density dependence of the hyperon couplings  to the $\sigma$,
$\omega$ and $\rho$ mesons are considered to be the same
 as one of the nucleons. For the $\phi$ coupling, the density
 dependence of  the $\omega$ meson is considered.
The couplings of the $\sigma$ meson to the $\Lambda$ and $\Xi$ have
 been taken from  \cite{Fortin:2017cvt} and \cite{Fortin:2020qin},
 respectively, and have been fitted to the binding energy of $\Lambda$
 and $\Xi$ hypernuclei. The coupling to the $\Sigma$ hyperon was
 chosen so that the $\Sigma$ potential in symmetric nuclear matter
 is $+30$ MeV, see \cite{Gal:2016boi} for a discussion.
 For the vector mesons, the quark model predictions are used,
\begin{align*}
&g_{\omega\Lambda}=g_{\omega\Sigma}=\frac{2}{3}g_{\omega N},\quad
g_{\omega\Xi}=\frac{1}{3}g_{\omega N},\\
&g_{\phi\Lambda}=g_{\phi\Sigma}=-\frac{\sqrt{2}}{3}g_{\omega N},\quad
g_{\phi\Xi}=-\frac{2\sqrt{2}}{3}g_{\omega N}.
\end{align*}
Finally,  the effective  $\rho$-meson coupling  is determined
by the product of the hyperon isospin with the $\rho$
meson-nucleon  coupling. The DD2 EoS without $\Lambda$ hyperons reproduces the maximum mass constraint $(M_{max}=2.4M_\odot)$ and NICER data,  while  $\Lambda_{1.4}$ is 700. At the same time, the DD2 EoS with $\Lambda$ hyperons gives the maximum mass of $2~M_{\odot}$.
Further on the DD2 EoS with $\Lambda$ hyperons will be referred as the DD2$\Lambda$.

The complete NS EoS contains,  besides the core EoS, the BPS EoS \cite{bps} for the outer crust, and the inner crust was calculated within a Thomas-Fermi calculation taking  DD2 as the underlying  model and allowing for the appearance of several geometries as discussed in \cite{grill14}. The inner crust EoS has been published in \cite{Fortin2016}. 

\section{Mixed system of two components}
\label{sec:MIXEoS}

We assume no interaction between DM and BM, except through gravity.
This assumption is fully justified by the latest constraints coming from the DM direct detection experiments and Bullet Cluster \citep{Clowe_2006, Randall:2008ppe}, showing that the DM-BM cross section is many orders of magnitude lower than the typical nuclear one,  $\sigma_\chi\sim 10^{-45}\ \mathrm{cm}^2\ll \sigma_N\sim10^{-24}\ \mathrm{cm}^2$.

Therefore, the stress-energy tensors of both components are conserved separately, leading to the system of the TOV equations with split components \citep{PhysRev.55.374,PhysRev.55.364}

\begin{equation}\label{TOV}
\frac{dp_i}{dr}=-\frac{(\epsilon_i +p_i)(M_\mathrm{tot}+4\pi r^3p_\mathrm{tot})}{r^2\left(1-{2M_\mathrm{tot}}/{r}\right)},
\end{equation}
which describes the relativistic hydrostatic equilibrium of a DM-admixed NS. In Eq.~\eqref{TOV}, the subscript index refers both to the BM and DM, i.e., $i=B,D$, while $p_{\mathrm{tot}}\equiv p_B+p_\chi$ and $M(r)$ are the total pressure and gravitational mass enclosed inside a sphere of radius $r$, respectively,
\begin{equation}\label{M_i}
    M_i(r) = 4\pi\int^r_0 \varepsilon_i (r^\prime)r^{\prime 2}dr^\prime.
\end{equation}

Using Eq.~\eqref{M_i}, we define the total gravitational mass as the sum of the two components, $M_\mathrm{tot} = M_B(R_B)+M_D(R_D)$, where the radii $R_i$ are evaluated using the zero-pressure condition at the surface
\begin{equation}
    p_i(R_i)=0.
\end{equation}

After having the total mass of the system, it is possible and convenient to write the fraction of the accumulated DM as
\begin{equation}\label{DMFRAC}
    f_\chi = \frac{M_D}{M_\mathrm{tot}}.
\end{equation}
It is worth noting, that we refer to the microscopic/thermodynamic DM parameters as $\chi$, while the macroscopic ones have an index $D$.

It is easy to obtain directly from Eq.~\eqref{TOV} the relation between the chemical potentials of the BM and DM. In fact, \cite{PhysRevD.102.063028} showed that
\begin{equation}
    \frac{d \ln \mu_B}{dr}=\frac{d \ln \mu_\chi}{dr} = -\frac{M_\mathrm{tot}+4\pi r^3 p_\mathrm{tot}}{r^2(1-2M_\mathrm{tot}/r)},
\end{equation}
which yields the conclusion that the two chemical potentials are proportional to each other. The value their ratio attains in the center of the star is the proportionality constant, which can be used to simplify the model
\begin{equation}\label{DMBM}
    \mu_\chi = \left(\frac{\mu_\chi}{\mu_B} \right)_{r=0} \mu_B.
\end{equation}

   \begin{figure*}
    \centering
    \setkeys{Gin}{width=1.15\linewidth}
    \begin{tabularx}{\linewidth}{XXX}
\includegraphics{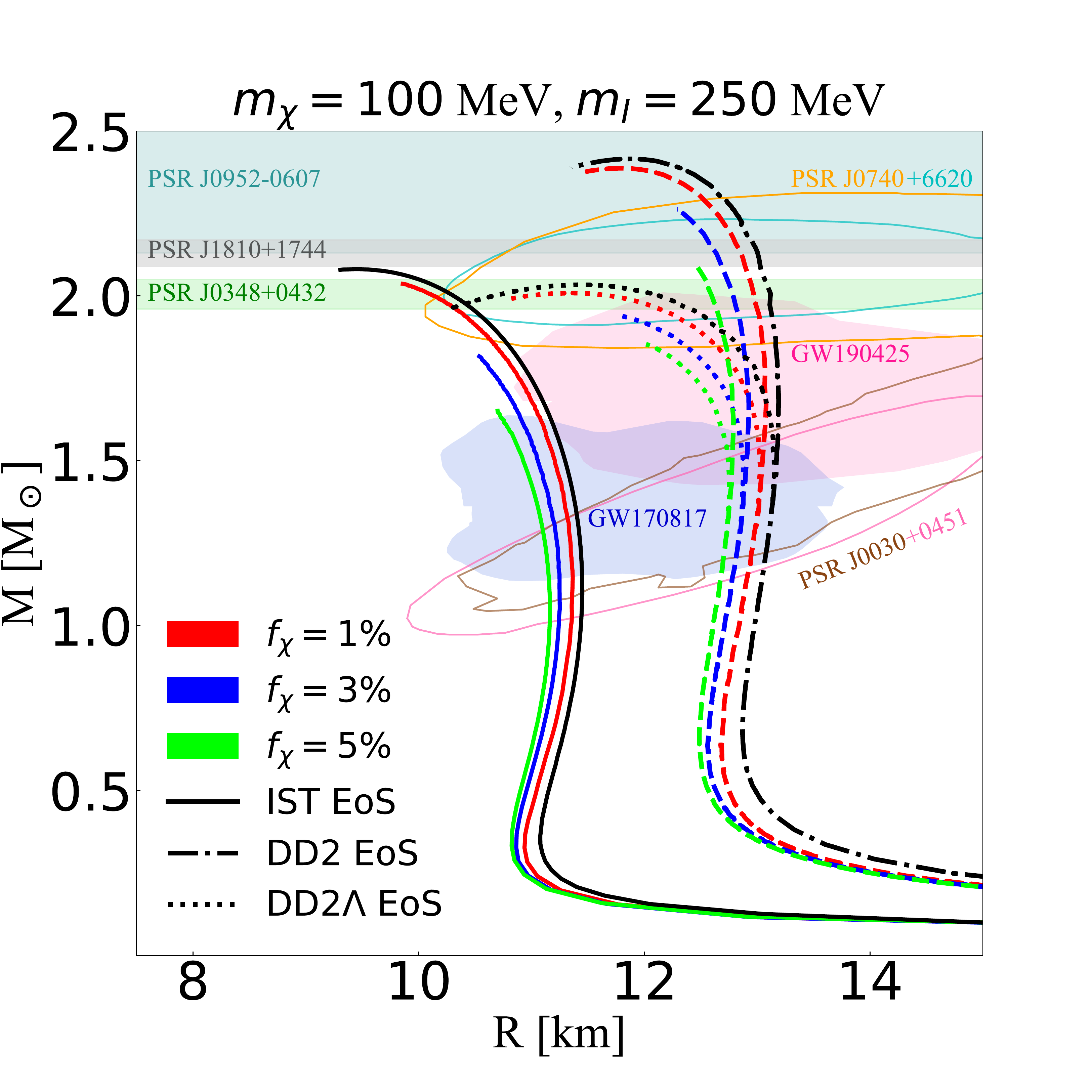}
    &
\includegraphics{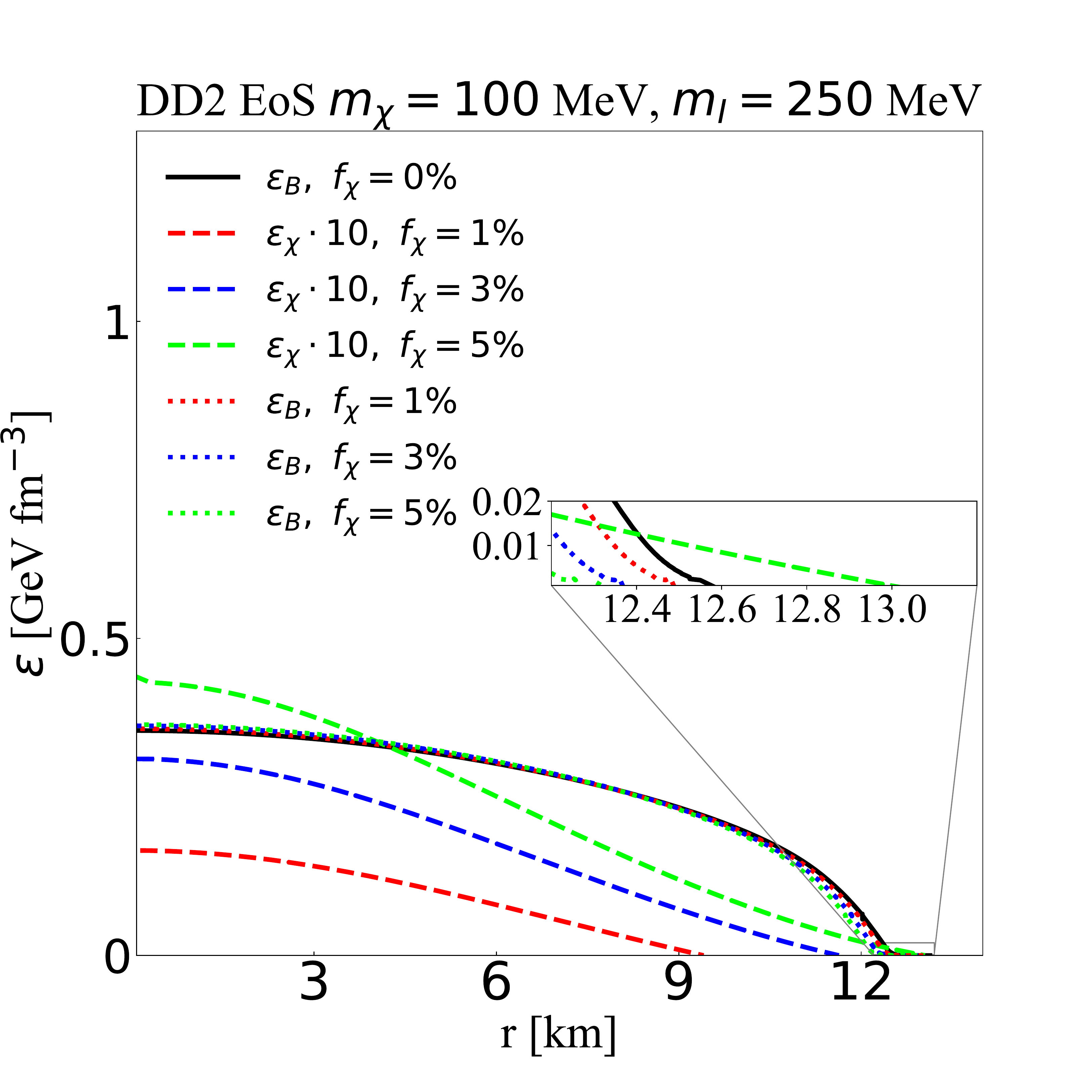}
    &
\includegraphics{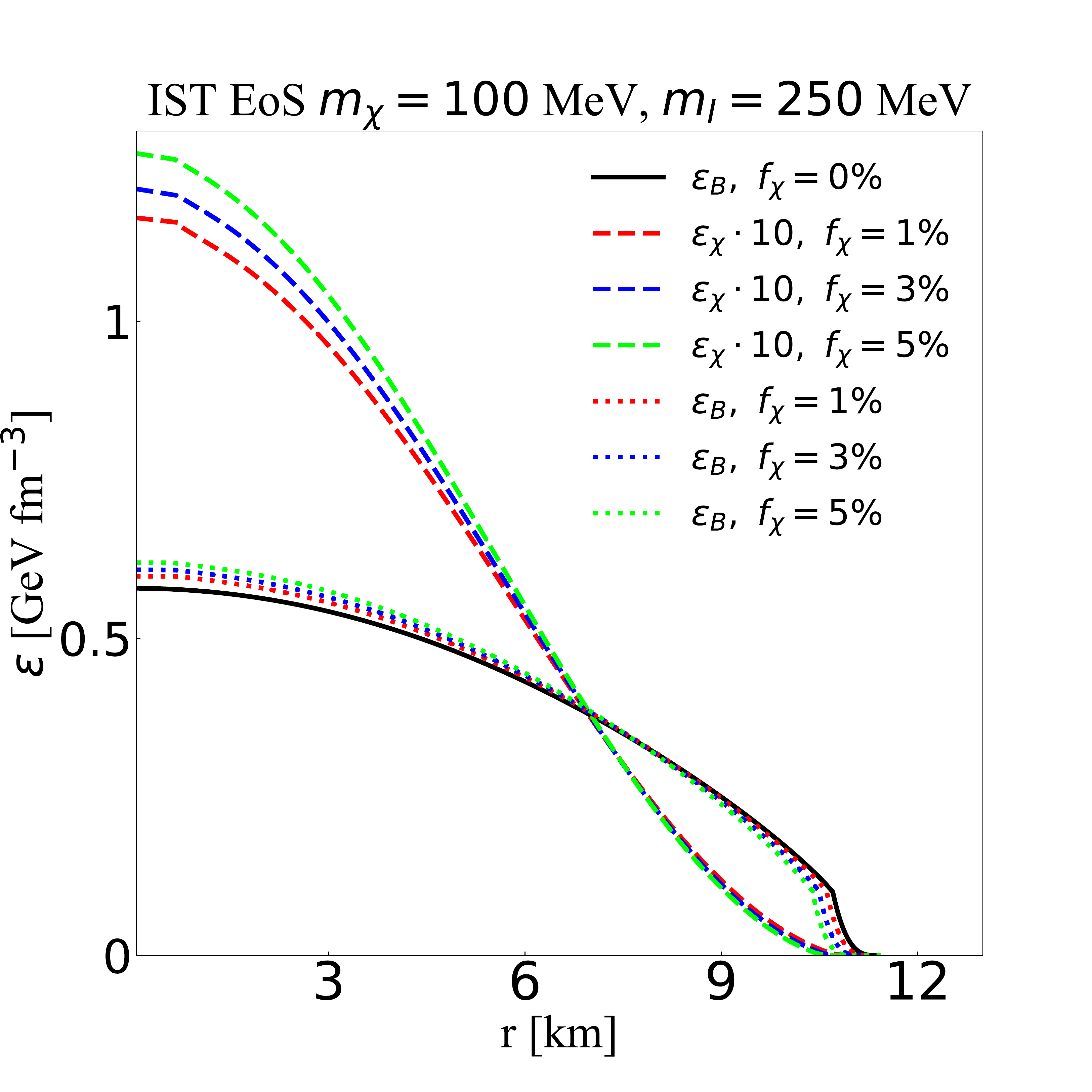}
    \end{tabularx}
        \begin{tabularx}{\linewidth}{XXX}
\includegraphics{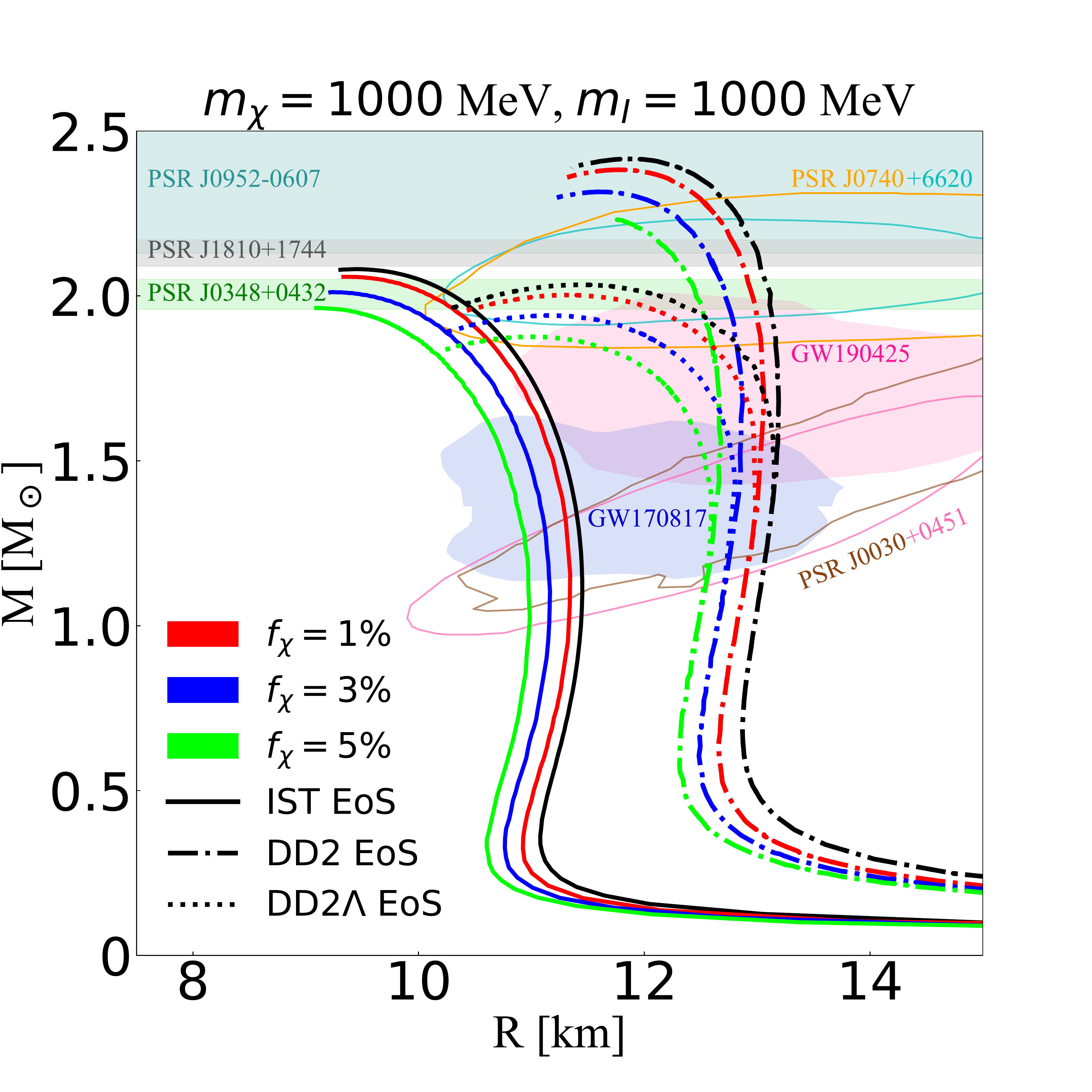}
    &
\includegraphics{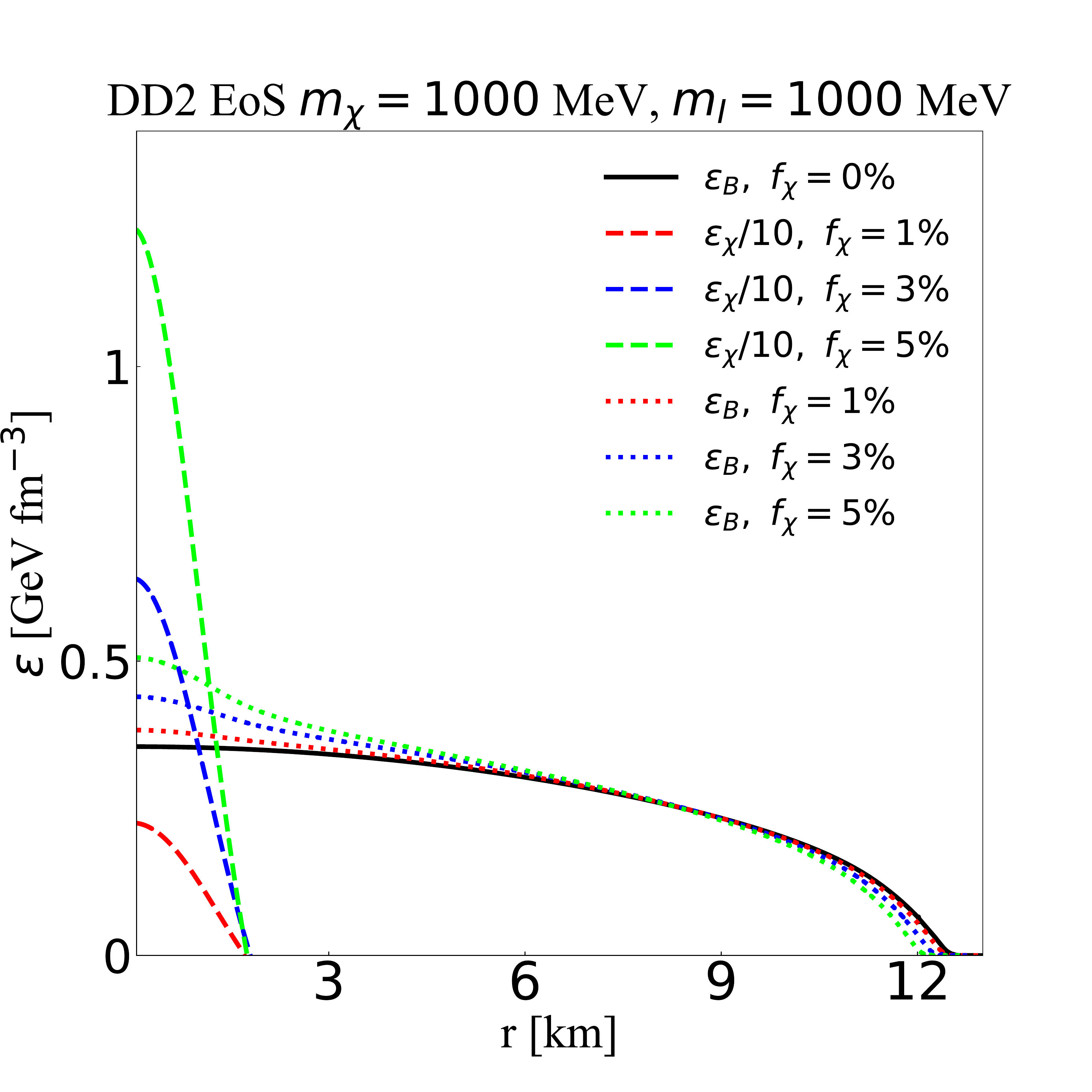}
    &
\includegraphics{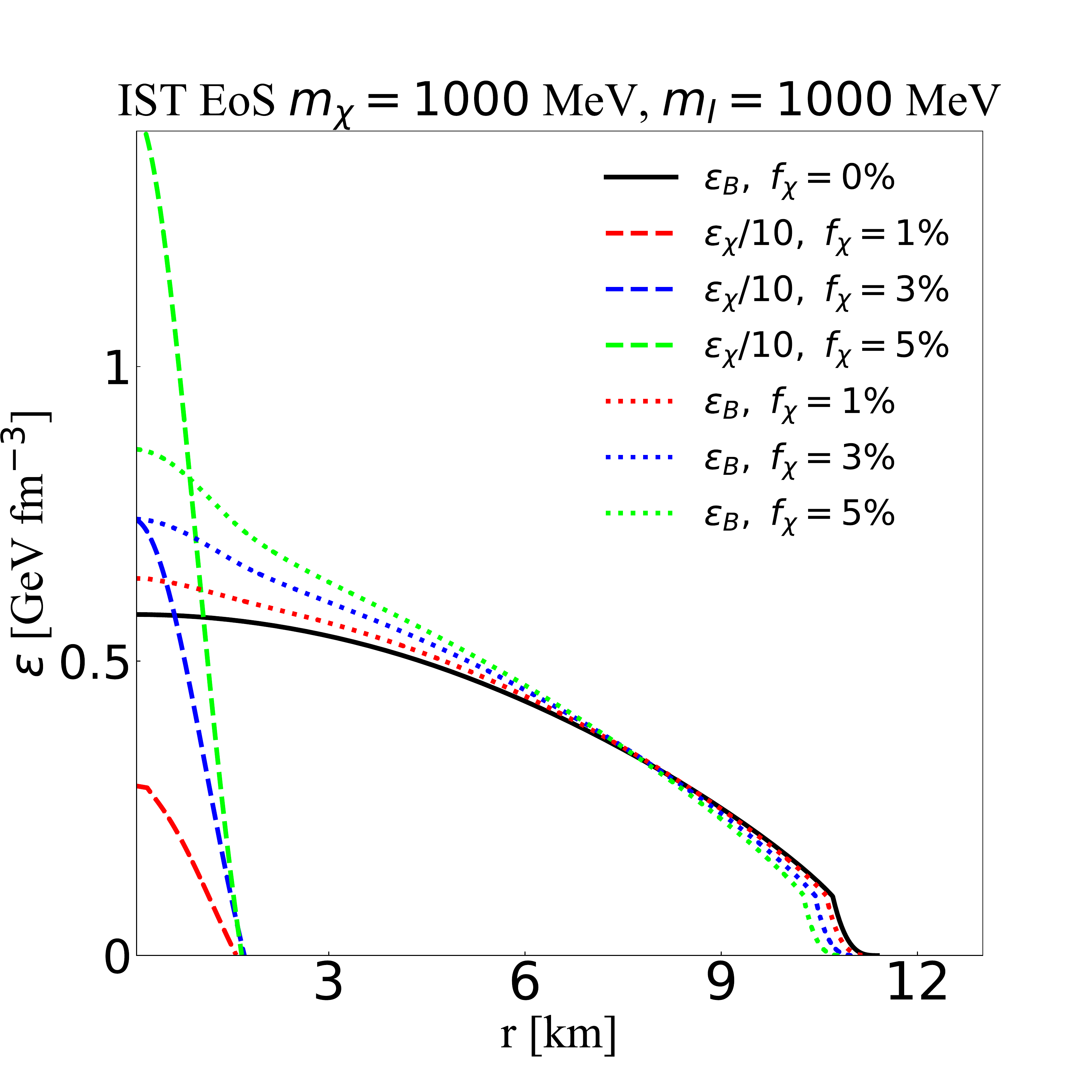}
    \end{tabularx}
    \caption{ Left column: total gravitational mass of the DM-admixed NS as a function of its visible radius $R$ obtained for $m_{\chi}$=100 MeV, $m_{I}$=250 MeV (upper panel) and $m_{\chi}$=1000 MeV, $m_{I}$=1000 MeV (lower panel). Black solid, dashed-dotted, and dotted curves correspond to pure BM stars described by the IST EoS, DD2 EoS, and DD2 EoS with hyperons. Red, blue, and green colours depict relative DM fractions equal to 1\%, 3\%, and 5\%, correspondingly. Green, gray, and teal bands represent 1$\sigma$ constraints on mass of PSR J0348+0432 \citep{PSRj03480432Article}, PSR J1810+1744 \citep{Romani:2021xmb}, and PSR J0952-0607 \citep{Romani:2022jhd}. Pink and beige contours show the NICER measurements of PSR J0030+0451 \citep{Miller_2019,Riley:2019yda}, while orange and blue contours depict the PSR J0740+6620 measurements \citep{Miller:2021qha,Riley:2021pdl}. LIGO-Virgo observations of GW170817 \citep{Abbott_2018} and GW190425 \citep{LIGOScientific:2020aai} binary NS mergers are shown in blue and magenta.
    Middle column: energy density profiles for the BM (dotted curves) and DM (dashed curves) components are shown for the DD2 EoS. The solid black curve represents the profile for pure BM $1.4~M_\odot$ NS, while the other profiles were sampled to have the same total gravitational mass. The upper panel is obtained for $m_{\chi}$=100 MeV, $m_{I}$=250 MeV and the lower one for $m_{\chi}$=1000 MeV, $m_{I}$=1000 MeV.
    Right column: the same as on the middle column, but for the IST EoS.
    }
\label{fig:MRprofiles}
    \end{figure*}

By solving the TOV Eq.~\eqref{TOV} with the boundary conditions and accounting for the relation between both components from Eq.~\eqref{DMBM}, we calculate the M-R relations for DM-admixed NSs for different values of DM fractions $f_\chi$, particle's mass $m_{\chi}$, and the interaction scale $m_{I}$. To better understand the impact of each parameter we consider light and heavy DM particles with $m_{\chi}$=100 MeV and $m_{\chi}$=1000 MeV (see the left column of Fig.~\ref{fig:MRprofiles}). Moreover, to address our ignorance of the EoS for baryonic component we studied the effect of DM on the soft IST EoS, depicted as a solid black curve on the left panels of Fig.~\ref{fig:MRprofiles}, as well as on the stiff DD2$\Lambda$ EoS (dotted black curve) and DD2 EoS (dashed-dotted black curve). The chosen EoSs represent different sides of mass and radius region allowed by the recent astrophysical, GW, and nuclear physics constraints, and therefore, provide good coverage of BM parameters. As it can be seen, the DD2$\Lambda$ EoS (dotted black curve) and DD2 EoS coincide until $\sim 1.4~M_{\odot}$, a point where the onset of hyperons happens. Further, hyperon production softens the EoS leading to a smaller total maximum mass and star's radius. 

The left panels of Fig.~\ref{fig:MRprofiles} show the effect of DM with different relative fractions inside a star on its mass and radius. Thus, we see a reduction of $M_{\rm max}$ and radius of stars for larger DM fractions caused by a DM core formation. In fact, the formation of more compact objects for an outside observer would look like a softening of the BM EoS. This degeneracy between the effect of DM and possible change of the strongly interacting matter properties at high density will be discussed in Section \ref{sec:Discussions}.

Due to the fact that in the considered model at $\mu_\chi \rightarrow \sqrt{2}m_\chi$ energy density diverges at finite pressure, DM falls under the Schwarzschild radius forming a black hole. It takes place for the high-mass stars for which the DM chemical potential in the center reaches the limit (see the upper left panel of Fig.~\ref{fig:MRprofiles}).

The panels on the middle and right columns of Fig.~\ref{fig:MRprofiles} demonstrate the split energy density profiles of DM (dashed curves) and BM (dotted curves). The solid black curve depicts the energy density profile for the $1.4~M_\odot$ star. The profiles for DM-admixed NSs are shown for stars with the same total gravitational mass as the pure BM NS. As the onset of hyperons occurs after $1.4 M_\odot$, two formulations of the DD2 EoS give the same prediction for the matter distribution inside the stars. Therefore, in Fig.~\ref{fig:MRprofiles} we show profiles only for the DD2 EoS. 

For heavy bosons a compact DM core is formed, which is seen from the high values of the $\epsilon_D$, being an order of magnitude above $\epsilon_B$ (see the middle and right panels of the low row isn Fig.~\ref{fig:MRprofiles}). Furthermore, the $\epsilon_D$ drops to zero at radius $\sim$2 km corresponding to the size of a DM core.

For the DM fraction 5\% and $m_{\chi}=100$ MeV, $m_{I}=$ 250 MeV (see the middle upper panel of Fig.~\ref{fig:MRprofiles}) a DM halo is formed with the radius of 13.0 km.

\section{Tidal deformability of DM-admixed NSs}
\label{sec:TID}

The tidal deformability parameter $\lambda$ quantifies the response of an object to a static external quadrupolar tidal field $\mathcal{E}_{ij}$ by relating it to a quadrupolar moment $\mathcal{Q}_{ij} = -\lambda\mathcal{E}_{ij}$. For a given stellar configuration of the total mass $M_{\mathrm{tot}}$ and radius $R$ this tidal deformability can be expressed through the Love number $k_2$ as $\lambda=2k_2 R^{5}/3$ and is commonly mapped to the dimensionless $\Lambda = \lambda/M_{\mathrm{tot}}^5$ \citep{Hinderer_2008}. In the two-component case, $R$ should be understood as the outermost radius, i.e., $R=R_{B}$ in the DM core scenario and  $R=R_{D}$ in the DM halo one. The Love number is defined through the solution of an ordinary differential equation (ODE) appearing as a leading order expansion of the Einstein equations with a metric perturbed by the external gravitational field \citep{1957PhRv..108.1063R}. The microscopic properties of matter are encoded into this ODE through the change of total pressure $p_{\mathrm{tot}}\equiv p_B+p_\chi$ caused by perturbation of the total energy density $\varepsilon_\mathrm{tot}\equiv \varepsilon_B+\varepsilon_\chi$. This change is quantified by the derivative $dp_{\mathrm{tot}}/d\varepsilon_{\mathrm{tot}}$. In the barotropic one-fluid case, this derivative represents the corresponding speed of sound. In the two-fluid case, the speed of sound derivation as $dp_\mathrm{tot}/d\varepsilon_\mathrm{tot}$ is mathematically identical to the expression obtained by \citet{Das:2020ecp}. Therefore, in what follows, we refer to it as the effective speed of sound of the two-fluid system. It can be expressed through the speed of sound of baryonic $c_{s,B}^2$ and dark $c_{s,\chi}^2$ components as
\begin{eqnarray}
\label{IX}
c_{s,\mathrm{eff}}^2=\eta c_\mathrm{s,B}^2+(1-\eta)c_{s,\chi}^2
\end{eqnarray}
with $\eta\in[0,1]$. The lower and upper edges of this interval correspond to the cases of pure DM and BM, respectively. Appendix \ref{appB:SpeedOfSound} provides the derivation of Eq.~\eqref{IX} and parameter $\eta$. This expression demonstrates that the effective speed of sound lies between the ones of pure components. 

\begin{figure}[t]
    \centering
    \includegraphics[width=0.95\columnwidth]{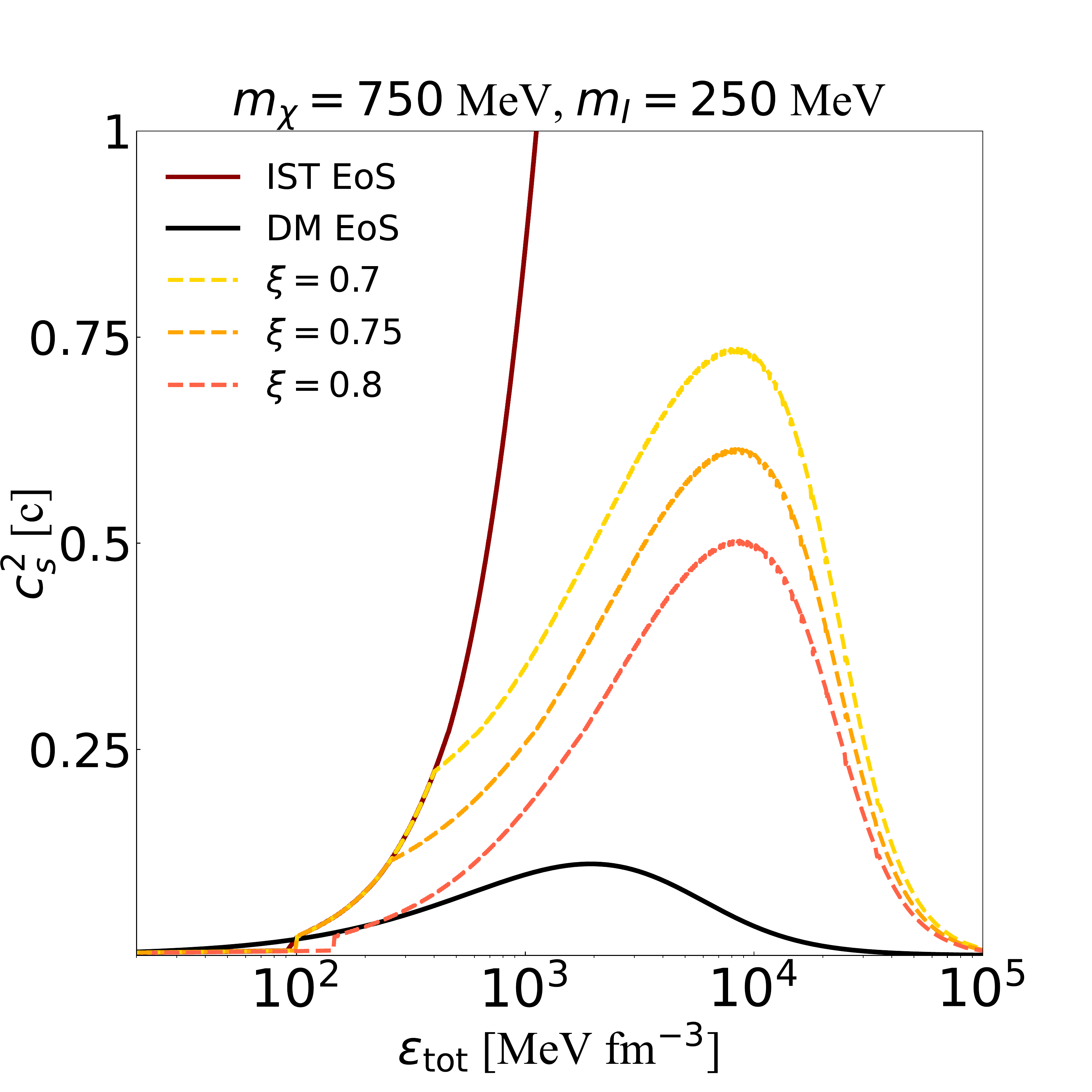}
    \includegraphics[width=0.95\columnwidth]{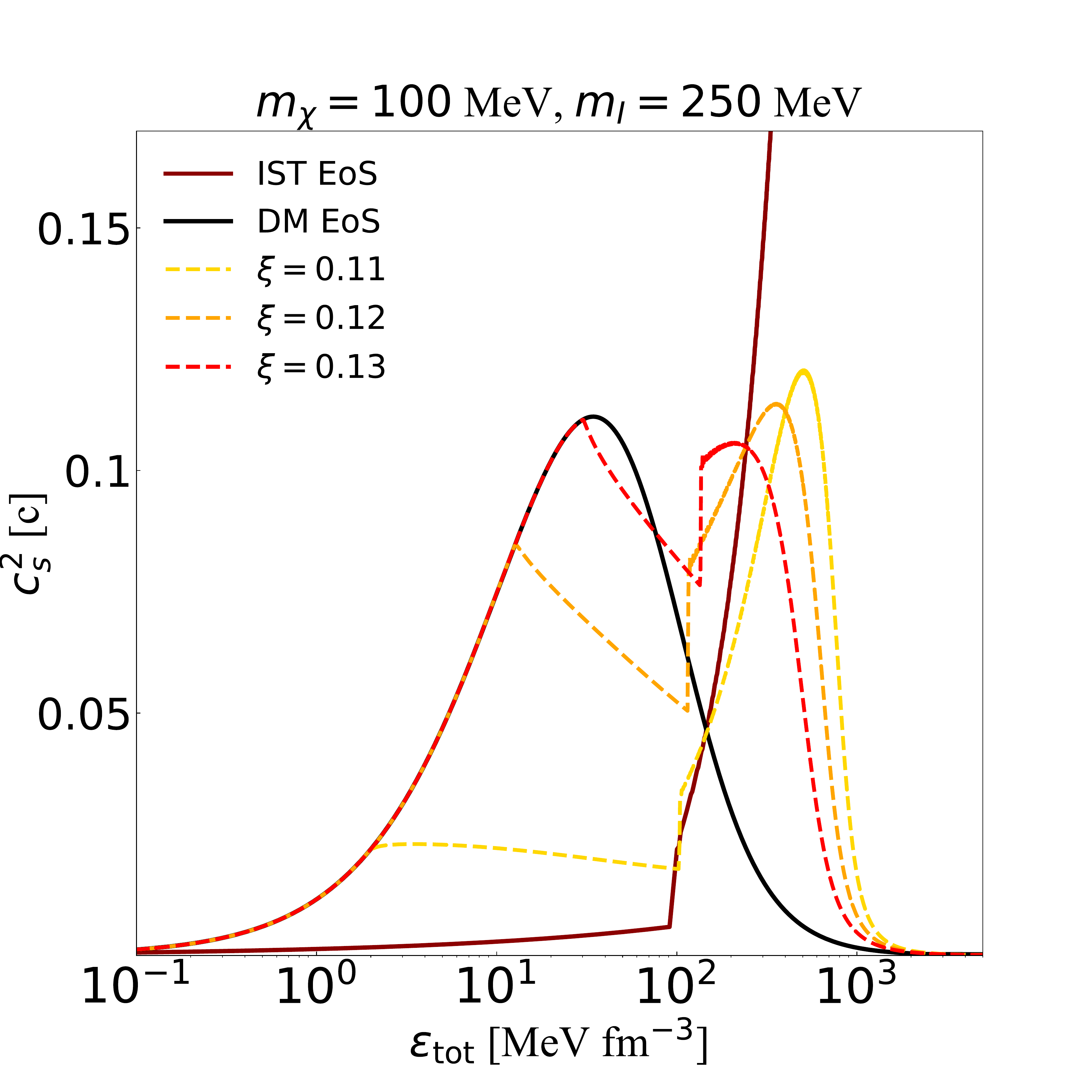}
    \caption{The effective speed of sound for a mixture of BM and DM as a function of total energy density. Upper panel: the curves were obtained for 
    $m_{\chi}$=750 MeV and $m_{I}$=250 MeV, which represents a DM core configuration. Lower panel: the same as on the upper panel, but for $m_{\chi}$=100 MeV and $m_{I}$=250 MeV illustrating a DM halo configuration. The horizontal line at low densities corresponds to the polytropic EoS for the crust.}
    \label{fig:sound}
\end{figure}

In Fig.~\ref{fig:sound} we show the effective speed of sound for different $\xi=\frac{\mu_\chi}{\mu_B}$ values, as well as the speed of sound for pure BM and DM components. A relation between the parameters $\xi$ and $\eta$ is given in Eq.~\eqref{B3} in Appendix \ref{appB:SpeedOfSound}. The upper panel of Fig.~\ref{fig:sound} indicates how the effective speed of sound behaves with DM accumulated in a core of a compact star. Note, that it is in between the speed of sound values for pure components. On the lower panel of Fig.~\ref{fig:sound} we see that the effective speed of sound follows the BM, and only in the outer crust the DM component stars dominate, which is related to a halo configuration.

\begin{figure}[t]
    \centering
    \includegraphics[width=0.95\columnwidth]{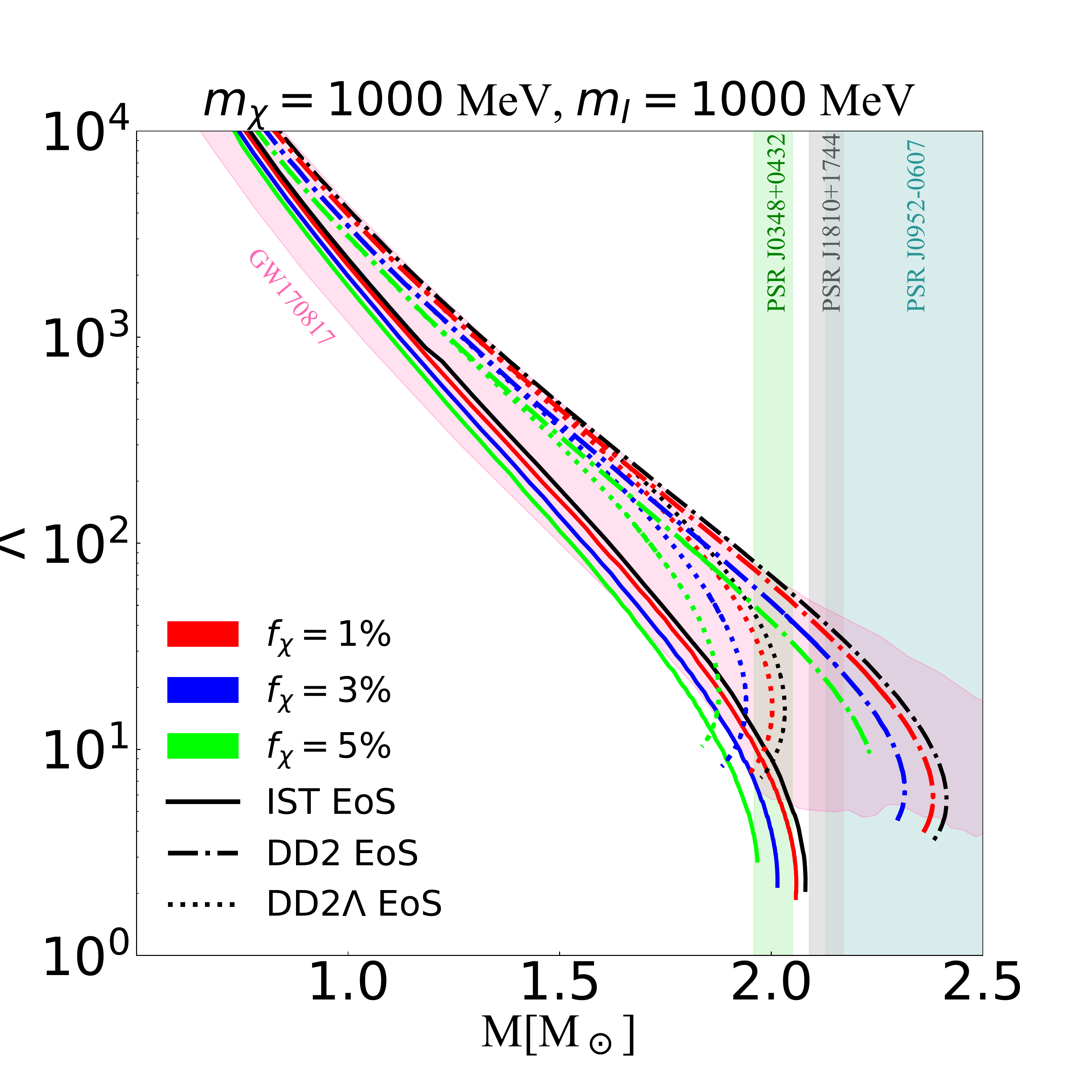}
    \caption{Tidal deformability as a function of total gravitational mass calculated for pure BM stars (black curves) and DM-admixed NSs with relative DM fractions 1\%, 3\%, and 5\%, in red, blue, and green, correspondingly. Solid, dash-dotted and dotted curves represent the IST EoS, DD2 EoS, and DD2$\Lambda$ EoS. The colors and symbols coincide with the ones used in Fig.~
    \ref{fig:MRprofiles} for a better comparison. The figure is obtained for $m_{\chi}$=1000 MeV, $m_{I}$=1000 MeV. Green, gray, and teal bands represent 1$\sigma$ constraints on mass of PSR J0348+0432 \citep{PSRj03480432Article}, PSR J1810+1744 \citep{Romani:2021xmb}, and PSR J0952-0607 \citep{Romani:2022jhd}. The magenta area visualizes the constraints obtained from GW170817 \citep{Abbott_2018}.}
    \label{fig:tides}
\end{figure}

As can be seen in Fig.~\ref{fig:tides}, for the given total gravitational mass DM condensed in a core leads to a smaller tidal deformability parameter compared to a pure baryonic star. A similar effect has been shown in Fig.~\ref{fig:MRprofiles} for the radius. For a distant observer, these effects  will be perceived as an effective softening of the EoS. On the other hand, the presence of a DM halo leads to a significant increase in the outermost radius that goes beyond the BM component, an increase of the tidal deformability parameter, and consequent effective stiffening of the EoS. The considered IST, DD2, and DD2$\Lambda$ EoSs make us conclude that the soft EoS, being on the lower limit of the GW170817 90\% CL region (see the magenta area in Fig.~\ref{fig:tides}), provides a stringent constraint on a DM core scenario, while the stiff EoS, being on the upper border of it, allows much higher DM fractions, and disfavors an extended halo configuration. This degeneracy between the effect of DM and strongly interacting matter properties at high densities possesses limitations on DM detection, except for several DM smoking guns that are going to be discussed in Section \ref{sec:Discussions}.
Despite it, we have to be aware of the fact that observational data on compact stars could be affected by accumulated DM and, consequently, constraints we put on strongly interacting matter at high densities.

\section{Results}
\label{sec:Results}

To study an interplay between boson mass and the interaction scale, as well as to put constraints on the DM fraction, we perform a scan over those parameters for the IST EoS (upper row), DD2 EoS (middle row), and DD2$\Lambda$ EoS (bottom row) for fixed DM fractions of 1\%, 3\%, and 5\% (see Fig.~\ref{fig:contours} in Appendix \ref{appC:Scan}). The color maps represent the total maximum gravitational mass of DM-admixed NSs. The white curve on each panel corresponds to $M_{max}=1.4~M_{\odot}$, whereas the red curve represents $M_{max}=2.0~M_{\odot}$. In the case $2.0~M_{\odot}$ configurations are not reachable, we indicate $1.9~M_{\odot}$ stars with a green curve. As one can see from the upper row on Fig.~\ref{fig:contours}, the increase of the DM fraction narrows the range of the values of the interaction scale $m_{I}$ consistent with the masses of the heaviest known pulsars. On the other hand, the existence of the high-mass stars with a significant amount of heavy DM requires low values of the interaction scale. 

We see the same dependence between $m_{\chi}$ and $m_{I}$ values. 
In fact, lower $m_I\equiv m_\omega/g$ values correspond to the higher coupling constant $g$ or, equivalently, stronger repulsion between the DM particles. The IST EoS for any DM fraction is always in agreement with the tidal deformability constraint, independently of $m_{\chi}$ and $m_{I}$ (see the upper row in Fig.~\ref{fig:contours}). At the same time, only for 1\% and 3\% of DM the total maximum mass of DM-admixed NSs can reach $2.0~M_{\odot}$. Thus, to simultaneously reproduce 
$2.0~M_{\odot}$ and GW170817 tidal deformability constraints the boson mass and interaction scale are restricted to the values shown in yellow. The shaded areas correspond to the non-allowed regions of parameters that cannot simultaneously provide the heaviest pulsars and GW constraints.

For 3\% and 5\% of DM the DD2 EoS reproduces both constraints in a wide range of parameters disfavouring megaelectronvolt mass range of bosonic DM with low values of the interaction strength. The black curve in the middle and bottom rows of Fig.~\ref{fig:contours} depicts the GW170817 tidal deformability constraint $\tilde{\Lambda}_{1.36} = 720$ \citep{LIGOScientific:2018hze} above which the model is consistent with the GW170817 merger. The dashed area corresponds to a non-allowed range of parameters, including 1\% of DM for the DD2 EoS. For the DD2$\Lambda$ EoSs there are no $m_{I}$ and $m_{\chi}$ values  that simultaneously reproduce the heaviest pulsars and GW constraints. In fact, only one of these criteria was reproduced for considered values of DM fractions. This is directly related to the fact that at the onset of $\Lambda$ hyperons the EoS becomes softer in addition to the DM softening effect in a core configuration.

From this analysis, we can conclude that, contrary to the stiff BM EoS (the DD2 EoS, as an example), the soft BM EoS (the IST EoS, as an example) provides a weaker limit on DM particle mass and interaction strength. This is related to the fact that the pure baryonic DD2 or DD2$\Lambda$ EoSs are on the upper border of the $\Lambda_{1.4}$ constraint from GW170817. Any decrease in $\Lambda_{1.4}$ due to a DM core will not violate this condition, whereas a small DM halo configuration will do it. As can be seen in Fig.~\ref{fig:tides}, the IST EoS is located on the lower limit of the magenta area favoring a halo formation. 
 
It is worth noting that this result is obtained under the assumption of a similar DM fraction in all galaxies. As a matter of fact, an application of the GW170817 tidal deformability result and multi-messenger data as a universal constraint on the amount of DM is questionable.
Each galaxy could be characterized by a different DM profile, as well as have local DM inhomogeneities. Strictly speaking, GW170817 probes an amount of DM only in a part of the NGC 4993, the host galaxy for this particular merger. Therefore, a larger sample size of NS-NS and NS-BH mergers is required to constrain the DM properties.

Due to current uncertainties of the BM EoS at high density, we cannot discriminate between the effect of DM and the properties of BM. As it will be discussed in the following Section \ref{sec:Discussions}, we expect a higher DM fraction inside compact stars toward the Galactic Center. If so, the compact star population would follow the scenarios presented from the left to right panels on Fig.~\ref{fig:contours}, i.e., from low to high DM fraction. 
\vspace*{1.15cm}
\section{Discussions}
\label{sec:Discussions}

As described above, there are various effects of DM on compact stars. A natural question arises: how we can narrow down the proposed DM models and constrain the DM properties using NSs? Can compact stars provide a smoking gun evidence for the presence of DM? There are several different approaches:

(i) By measuring the mass, radius, and moment of inertia of NSs with few-percent accuracy. Nowadays, NICER \citep{Miller_2019,Raaijmakers:2019dks,Miller:2021qha,Raaijmakers:2021uju} and in the near future ATHENA \citep{Cassano:2018zwm}, eXTP \citep{eXTP:2018kws}, and STROBE-X \citep{STROBE-XScienceWorkingGroup:2019cyd} are expected to measure $M$ and $R$ of NSs with a high accuracy. Using the synthetic data for the STROBE-X telescope, and assuming two NSs of the same mass and BM EoS, \citet{Rutherford:2022xeb} concluded that a measurement of radii with a 2\% accuracy would be enough to draw a conclusion about the presence of DM in star's interior. However, the existence of the deconfinement phase transition in a core would exhibit in the same way, leading to a degeneracy between the effect of DM and the phase transition. The main drawback of this approach is that the effect of DM could mimic the softening/stiffening of BM at high density and vice versa. Current uncertainties of the baryonic EoS do not allow discrimination of two effects.
In addition, radio telescopes, e.g., MeerKAT \citep{Bailes:2018azh}, SKA \citep{Watts:2014tja}, and ngVLA \citep{Bower:2018mta} plan to increase radio pulsar timing and discover Galactic Center pulsars. A mass reduction of NSs toward the Galactic Center or variation of mass, radius, and moment of inertia in different parts of the Galaxy could shed light on the amount of accumulated DM in compact stars. In fact, we could see a paucity of old millisecond pulsars in the Galactic Center either due to light extinction on dust, or the collapse of DM-admixed NSs into black holes after exceeding the Schwarzschild limit  \citep{Bramante:2014zca}.

(ii) By performing binary numerical-relativity simulations and kilonova ejecta for DM-admixed compact stars for different DM candidates, mass of particles, interaction strength, and fractions with the further comparison to GW and electromagnetic signals. The smoking gun of the presence of DM could be a supplementary peak in the characteristic GW spectrum of NS mergers \citep{Ellis:2017jgp}, exotic waveforms \citep{Giudice:2016zpa}, modification of the kilonova ejecta, or the presence of a strong oscillation mode in the waveforms during the post-merger stage \citep{Bezares:2019jcb}. The next generation of GW detectors, i.e., the Cosmic Explorer (CE) \citep{Mills:2017urp} and Einstein Telescope (ET) \citep{Punturo:2010zz} will open another perspective of detection of post-merger regimes and probing an internal composition of compact stars.

(iii) By detecting a new feature in the binary Love relation 
\citep{Yagi:2015pkc}. Thus, as it was shown in Fig.~\ref{fig:sound}, DM could produce a bump, or any other irregular behavior, in the effective speed of sound that would affect the binary Love relation. Similar, as it was demonstrated for the strongly interacting matter by \citet{Tan:2021nat}. 
This mark may be revealed by the next generation of GW detectors that are planned to have the measurement precision of $\delta \Lambda \sim 5$ for a GW170817-like event.

(iv) By detecting objects that go in contradiction with our understanding. A potential candidate for DM-admixed NS could be the secondary component of GW190814 \citep{LIGOScientific:2020zkf}. While likely being a black hole \citep{Essick:2020ghc,Tews:2020ylw}, this compact object with the mass of $\sim2.6~M_{\odot}$ raised debates about its nature \citep{Tsokaros:2020hli} as a pure baryon matter EoS would not be able to explain a compact star of $\sim2.6~M_{\odot}$. Hence, if not being a black hole, the compact object would have to be supplemented either with exotic degrees of freedom, such as hyperons and/or quarks \citep{Tan:2020ics,Dexheimer:2020rlp}, an early deconfinement phase transition \citep{Ivanytskyi:2022oxv}, very fast rotation \citep{Zhang:2020zsc}, or extra stiffening of the EoS at high densities \citep{Fattoyev:2020cws}. An alternative explanation of this puzzle would be a DM-admixed NS \citep{DiGiovanni:2021ejn}, which could also explain a formation of a black hole of so low mass as a collapsed DM-admixed NS \citep{Bramante:2014zca}.

The recently announced measurement of the central compact object within the supernova remnant HESS J1731-347 \citep{Doroshenko:2022} is another object that puzzles our understanding. This lightest and smallest compact star ever observed could be explained as a NS admixed with DM ~\citep{2023Sagun}. 

(v) Modification of the pulsar pulse profile due to the extra light-bending \citep{Miao:2022rqj} and/or gravitational microlensing in the case of the existence of a dark halo.

(vi) Modification of the cooling rate of compact stars \citep{2010PhRvD..81l3521D,Hamaguchi:2019oev, AngelesPerez-Garcia:2022qzs,Buschmann:2021juv}. We want to note, that this effect is the most inaccurate among the abovementioned ones. Thus, NSs need to have a well-measured surface luminosity and age. In addition to it, uncertainties related to particle composition, EoS, magnetic field, superfluidity/superconductivity, NS masses, the chemical composition of an atmosphere, etc., could wash out an effect of DM. Old NSs are less affected by the mentioned effects, as a photon cooling stage starts to dominate over a neutrino cooling stage that is very sensitive to a particle composition and superfluidity/superconductivity \citep{Page:2004fy}. The magnetic field is also expected to be unimportant for old isolated NSs. Therefore, a possible heating mechanism of NSs due to DM annihilation could be probed by increasing statistics on observational data of old NSs.

\section{Conclusions}
\label{sec:Concl}

We proposed a model of bosonic DM represented by a complex scalar field coupled to the vector one through the covariant derivative, which is equivalent to scalar electrodynamics with massive photons. The model describes DM existing in the form of BEC with repulsive interaction. Pressure of the present EoS saturates at asymptotically high densities leading to the vanishing speed of sound and compressibility at this regime. From the thermodynamic requirements, the chemical potential of DM existing as such BEC is limited to the interval $\mu_\chi\in[m_\chi,\sqrt{2}m_\chi]$, with $m_\chi$ being the DM particle mass. In the weak and strong coupling limits, this interval shrinks to its lower and upper bounds, respectively, while pressure vanishes even at any density. This spectacular feature of the present model makes its weak and strong coupling limits qualitatively similar and requires further clarification.

At the same time, the strong coupling limit is shown to be inconsistent with the Bullet Cluster cosmological data. Confronting the model prediction on the cross section of the DM self-interaction to the results of numerical simulations and observations allowed us to constrain the interaction scale $m_I$ from below depending on the DM particle mass $m_\chi$.

DM-admixed compact stars were modeled by considering the mixed system of two fluids with different relative fractions. The performed derivation of the effective speed of sound for a two-fluid system allowed us to calculate the tidal deformability parameter for compact stars admixed with different amounts of DM. We argue that the one-fluid approach cannot be applied to a mixed system of several components with the different proper speed of sound values.

To account for a discrepancy related to the baryonic component the soft IST EoS and stiffer DD2 EoS with and without hyperons were considered. For different DM particle's mass, its relative fraction, and interaction scale we found the conditions of DM core formation. We argue that in the framework of the considered model only a small DM halo is possible, with the outermost radius around twice the baryonic one. However, the total maximum gravitational mass of this configuration is below 2~$M_{\odot}$.

We performed a thorough analysis of the effect of DM particle mass in the 100-1000 MeV mass range and self-interacting scale on maximum total gravitational mass and tidal deformabilities of NSs for several fixed DM fractions. We found that for 1\%, 3\% of DM for the IST EoS and 3\%, 5\% of DM for the DD2 EoS the model can simultaneously reproduce the heaviest pulsars and GW170817 tidal deformability constraint. The obtained allowed region of boson mass $m_{\chi}$ and interaction scale $m_{I}$ for a fixed DM fraction shows an anticorrelated dependence between these parameters, i.e., a high $m_{\chi}$ value favors a low $m_{I}$ value. For the DD2$\Lambda$ EoS no allowed region of parameters was found due to the inability to simultaneously reproduce both constraints. 

The considered DM fractions up to 5\% were chosen to demonstrate the effects on compact star properties, and as it was discussed in the Introduction, are higher than could be accumulated by Bondi accretion. While up-to-date calculations are based on the interaction between DM and BM, the self-interaction of DM could lead to enhanced DM accretion, hence the DM fraction here proposed. However, we leave this for future studies.
 
In Section \ref{sec:Discussions}, we discussed the possible smoking gun signatures of DM in compact stars that could be probed in the near future, e.g., alteration of maximum total gravitational mass and radius of compact stars as a function of a distance from the Galactic Center; modification of the surface temperature (an additional heating or cooling mechanism) of NSs towards the Galactic Center; lack of old millisecond pulsars in the Galactic center; the presence of supplementary peak(s) in the GW signal from NS-NS and/or NS-BH mergers, exotic waveform, or modification of the kilonova ejecta; gravitational-lensing effect or alteration of the pulsar pulse profile due to the extra light bending in a dark halo. Moreover, such objects as a secondary component of the GW190814 event and central compact object within the supernova remnant HESS J1731-347 challenge the existing models of compact stars and black holes, offering the possibility of these objects being a DM-admixed NS. 

We argue that compact stars and their mergers provide a novel sensitive indirect method of detection and constraining the DM properties. Based on the performed analysis it is clear that the present data analysis of X-ray, radio, and GW observations without accounting for an accumulated DM could miss a valuable piece of information as well as give an incorrect prediction about the strongly interacting matter properties at high density.

\begin{acknowledgments}
The work of E.G., C.P., and V.S. was supported by national funds from FCT – Fundação para a Ciência e a Tecnologia, I.P., within Projects No.  UIDB/04564/2020, UIDP/04564/2020, EXPL/FIS-AST/0735/2021. E.G. also acknowledges the support from Project No. PRT/BD/152267/2021. C.P. is supported by the Project No. PTDC/FIS-AST/28920/2017. V.S. also acknowledges the PHAROS COST Action CA16214. The work of O.I. was supported by the Polish National Science Center under the grant No. 2019/33/BST/03059.
\end{acknowledgments}

\appendix
\section{Lagrangian Model of the DM EOS}
\label{appA:FullLagrangian}

Here, we derive the DM EoS, presented with Eqs.~\eqref{EqI} and~\eqref{EqII}, and we analyze it in the weak and strong coupling regimes. The minimal Lagrangian describing the chosen model of DM should include mass and kinetic terms of the complex scalar $\chi$ and real vector $\omega^\mu$ fields, which are coupled through the covariant derivative $D^\mu=\partial^\mu-ig\omega^\mu$ with $g$ being the corresponding Yukawa coupling constant. Thus, we start with
\begin{equation}
\label{A1}
\mathcal{L} = (D_\mu\chi)^*D^\mu\chi-m_\chi^2 \chi^*\chi -\frac{\Omega_{\mu\nu}\Omega^{\mu\nu}}{4}+\frac{m_\omega^2\omega_\mu\omega^\mu}{2},
\end{equation}
where $m_\chi$ and $m_\omega$ are masses of the scalar and vector fields, respectively, and $\Omega_{\mu\nu}=\partial_\mu\omega_\nu-\partial_\nu\omega_\mu$. Before going further we would like to discuss the Neother current resulting from the Lagrangian~\eqref{A1}
\begin{equation}
\label{A2}    
j^\mu=i\left[\chi^*\frac{\partial\mathcal{L}}{\partial(\partial_\mu\chi^*)}-\chi\frac{\partial\mathcal{L}}{\partial(\partial_\mu\chi)}\right]=
i\left[\chi^*D^\mu\chi-\chi(D^\mu\chi)^*\right]=
i(\chi^*\partial^\mu\chi-\chi\partial^\mu\chi^*)+
2g\omega^\mu\chi^*\chi.
\end{equation}
This current is equivalent to the one corresponding to local U(1) symmetry of $\mathcal{L}$ \citep{Brading:2000hc}. The density of conserved charge associated to the DM particles is obtained by averaging the zeroth component of this current. The second term in the expression for $j^\mu$ vanishes under such averaging, since it includes an odd number of creation and annihilation operators of the vector field. Thus, 
$n_\chi=\langle j^0\rangle=i\langle\chi^*\partial^0\chi-\chi\partial^0\chi^*\rangle$. This relation is artificially violated in the case when $\omega^\mu$ is treated classically, e.g., within the mean-field approximation. This discrepancy can be removed by using the Noether current resulting from the invariance with respect to global \textit{U}(1) transformation \citep{Brading:2000hc}
\begin{equation}
\label{A3}    
j^\mu=i(\chi^*\partial^\mu\chi-\chi\partial^\mu\chi^*).
\end{equation}
Below we use this expression for the DM four current.

Within the mean-field approximation operator of the vector field in this Lagrangian is replaced by its constant expectation value. This value can be obtained from the corresponding Euler-Lagrange equation 

\begin{equation}
\label{A4}
(\partial^2+m_\omega^2+2g^2\chi^*\chi)\omega^\mu-
\partial^\mu\partial^\nu\omega_\nu+gj^\mu=0.
\end{equation}
In the hydrostatic case only the zeroth component of the DM four-current attains nonzero mean value being nothing else as the DM particle number density $n_\chi$, i.e., $\langle j^\mu\rangle=\eta^{\mu0}n_\chi$ with $\eta^{\mu\nu}$ standing for the Minkowski metric tensor. Replacing $\omega^\mu$ by its constant mean value we eliminate derivatives of the vector field in Eq.~\eqref{A4}. Furthermore, we replace $\chi^*\chi$ by the scalar field condensate $\langle\chi^*\chi\rangle$ and DM current by its mean value. Thus, the Euler-Lagrange equation of the vector field under the mean-field approximation can be given a form of
\begin{equation}
\label{A5}
\langle\omega^\mu\rangle=\eta^{\mu0}\omega
\quad{\rm with}\quad
\omega=-\frac{gn_\chi}{m_\omega^2+2g^2\langle\chi^*\chi\rangle}.
\end{equation}
At finite particle number densities, DM has finite chemical potentials $\mu_\chi$, which serves as a Lagrange multiplier in $\mathcal{L}+\mu_\chi j^0$ ensuring that mean value of $j^0$ coincides with $n_\chi$. Inserting Eq.~\eqref{A5} into Eq.~\eqref{A1}, one gets 
\begin{equation}
\label{A6}
\mathcal{L}+\mu_\chi j^0 = \partial_\mu\chi^*\partial^\mu\chi-M_\chi^2 \chi^*\chi+\nu_\chi j^0+\frac{m_\omega^2\omega^2}{2}.
\end{equation}
The first three terms on the right-hand side of this equation represent free quasi-particles with the effective mass and chemical potential defined as 
\begin{eqnarray}
 \label{A7}
M_\chi^2&\equiv& m_\chi^2-g^2\omega^2,\\
\label{A8}
\nu_\chi&\equiv&\mu_\chi+g\omega,
\end{eqnarray}
respectively. At zero temperature these bosonic quasi-particles condense to zero mode with $\langle\chi^*\chi\rangle=\zeta^2$ and $\zeta$ being an amplitude of this mode. This BEC contributes to the DM pressure $\zeta^2(\nu_\chi^2-M_\chi^2)$ (see, e.g., chapter 2.4 of \citet{kapusta_gale_2006}). The last term in Eq.~\eqref{A6} does not include any dynamical variables, and therefore, simply renormalizes the pressure, which becomes
\begin{equation}
 \label{A9}
 p_\chi=\zeta^2(\nu_\chi^2-M_\chi^2)+\frac{m_\omega^2\omega^2}{2}.
\end{equation}
Mean value of the vector field defined by Eq.~\eqref{A4} maximizes the pressure. The same is the case for $\zeta$, i.e., $\frac{\partial p_\chi}{\partial\omega}=\frac{\partial p_\chi}{\partial\zeta}=0$. These conditions should be supplemented with a definition of the DM particle number density given by the thermodynamic identity $n_\chi=\frac{\partial p_\chi}{\partial\mu_\chi}$ yielding the following system of equations, which should be solved in order to find $\omega$ and $\zeta$ and construct the DM EoS:
\begin{eqnarray}
\label{A10}
&&2g\zeta^2(\nu_\chi+g\omega)+m_\omega^2\omega=0,\\
\label{A11}
&&2\zeta(\nu_\chi^2-M_\chi^2)=0,\\
\label{A12}
&&n_\chi=2\zeta^2\nu_\chi.
\end{eqnarray}
Before solving this system, we want to demonstrate that it is consistent with Eq.~\eqref{A5}. For this we formally express $\nu_\chi$ from Eq.~\eqref{A12} and insert the result to Eq.~\eqref{A10} yielding $gn_\chi+(2g^2\zeta^2+m_\omega^2)\omega=0$. Then, replacing $\zeta^2$ by $\langle\chi^*\chi\rangle$ this relation can be written in the desired form.

In the BEC, amplitude of the bosonic zero mode is $\zeta\neq0$. Therefore,  Eq.~\eqref{A11} requires $\nu_\chi=M_\chi$, which can be solved with respect to the vector field as
\begin{equation}
\label{A13}
\omega = \frac{-\mu_\chi\pm\sqrt{2m_\chi^2-\mu_\chi^2}}{2g}.
\end{equation}
From this condition we immediately conclude that physical values of the DM chemical potential are limited to the range $-\sqrt{2}m_\chi\le\mu_\chi\le\sqrt{2}m_\chi$. At positive $\mu_\chi$ BEC is constituted by the DM particles, while negative $\mu_\chi$ corresponds to antiparticles. For definiteness in this work, we consider the case of the DM particles, which is equivalent to requiring $\mu_\chi\ge0$. Eq.~\eqref{A13} ensures that the first term in Eq.~\eqref{A9} vanishes and DM pressure $p_\chi\propto\omega^2$. At zero $n_\chi$ this pressure should vanish provided by $\omega=0$. At non-negative chemical potential of DM this condition can be fulfilled only at $\mu_\chi=m_\chi$ if the sign ``$+$'' is chosen in Eq.~\eqref{A13}. In order to obtain the DM particle number density we first express $\zeta^2$ from Eq.~\eqref{A10} and then insert the result into Eq.~\eqref{A12}. Thus, at $\mu_\chi\in[m_\chi,\sqrt{2}m_\chi]$ the DM EoS becomes
\begin{eqnarray}
\label{A14}
p_\chi&=&\frac{1}{4}\left(\frac{m_\omega}{g}\right)^2 
\left(m_\chi^2-\mu_\chi\sqrt{2m_\chi^2-\mu_\chi^2}\right),\\
\label{A15}
n_\chi&=&\frac{1}{2}\left(\frac{m_\omega}{g}\right)^2 
\frac{\mu_\chi^2-m_\chi^2}{\sqrt{2m_\chi^2-\mu_\chi^2}},
\end{eqnarray}
while at $\mu_\chi\in[0,m_\chi]$ one gets $p_\chi=0$ and $n_\chi=0$. It is seen from these expressions that vector field mass $m_\omega$ and coupling $g$ do not enter the DM EoS independently but appear as the ratio $m_I\equiv m_\omega/g$, which is a relevant parameter. With this notation and thermodynamic identity $\varepsilon_\chi=\mu_\chi n_\chi-p_\chi$ we arrive at Equations~\eqref{EqI} and~\eqref{EqII}.

The weak and strong coupling limits of the present EoS are obtained at $g\rightarrow0$ and 
$g\rightarrow\infty$, respectively. This corresponds to $m_I\rightarrow\infty$ and $m_I\rightarrow0$. In order to consider these limits we treat the DM particle density as an independent quantity. For this we first write Eq.~\eqref{A15} as a quadratic equation for $\sqrt{2m_\chi^2-\mu_\chi^2}$ and solve it as
\begin{eqnarray}
\label{A16}
\sqrt{2m_\chi^2-\mu_\chi^2}=
\sqrt{m_\chi^2+\frac{n_\chi^2}{m_I^4}}-
\frac{n_\chi}{m_I^2}.
\end{eqnarray}
This solution allows us to express $\mu_\chi$ and expand it up to the leading order in $m_I^{-2}$ or $m_I^2$
\begin{eqnarray}
\label{A17}
\mu_\chi=
\left\{
\begin{array}{l}
m_\chi+\frac{n_\chi}{m_I^2}+
\mathcal{O}\left(m_I^{-4}\right),\quad\quad~~~ m_I\rightarrow\infty\\
\sqrt{2}m_\chi-\frac{m_\chi^3m_I^4}{8\sqrt{2}n_\chi^2}+\mathcal{O}\left(m_I^{8}\right),\quad m_I\rightarrow0.
\end{array}
\right.
\end{eqnarray}
From this we conclude that in the weak coupling limit chemical potential of DM converges to the smallest value of the BEC interval $\mu_\chi\in[m_\chi,\sqrt{2}m_\chi]$, while in the strong coupling limit it converges to the largest value. This conclusion holds for any $n_\chi$. The DM pressure behaves as
\begin{eqnarray}
\label{A18}
p_\chi=\frac{m_I^2}{8}\left(\mu_\chi-\sqrt{2m_\chi^2-\mu_\chi^2}\right)^2=
\left\{
\begin{array}{l}
\frac{n_\chi^2}{2m_I^2}+
\mathcal{O}\left(m_I^{-4}\right),\quad m_I\rightarrow\infty\\
\frac{m_\chi^2m_I^2}{4}+\mathcal{O}\left(m_I^{4}\right),
\quad m_I\rightarrow0.
\end{array}
\right.
\end{eqnarray}
where on the second step $\sqrt{2m_\chi^2-\mu_\chi^2}$ was approximated by $m_\chi-n_\chi/m_I^2$ at $m_I\rightarrow\infty$, while at $m_I\rightarrow0$ it was neglected compared to $\mu_\chi$. In both of the considered limits the pressure vanishes leading to the energy density mentioned in Subsection \ref{subsec:DMEoS}.

\section{Effective Speed of Sound of Two-Fluid System}
\setcounter{equation}{0}
\label{appB:SpeedOfSound}

In order to calculate $c_{s,\chi}^2$ we notice that in the GCE pressure and energy density of each component are functions of the corresponding chemical potential. With this we can write
\begin{eqnarray}
\label{B1}
c_{s,tot}^2=\frac{dp_{tot}}{d\varepsilon_{tot}}=
\frac{\frac{\partial p_B}{\partial\mu_B}+
\frac{\partial p_\chi}{\partial\mu_\chi}\frac{d \mu_\chi}{d\mu_B}}
{\frac{\partial \varepsilon_B}{\partial\mu_B}+
\frac{\partial \varepsilon_\chi}{\partial\mu_\chi}\frac{d \mu_\chi}{d\mu_B}}=
\frac{\frac{\partial \varepsilon_B}{\partial\mu_B}c_{s,B}^2+
\frac{\partial \varepsilon_\chi}{\partial\mu_\chi}\frac{d\mu_\chi}{d\mu_B}c_{s,\chi}^2}
{\frac{\partial \varepsilon_B}{\partial\mu_B}+
\frac{\partial \varepsilon_\chi}{\partial\mu_\chi}\frac{d \mu_\chi}{d\mu_B}},
\end{eqnarray}
where on the last step we used identities $c_{s,B}^2=\frac{\partial p_B}{\partial\mu_B}\bigl/\frac{\partial \varepsilon_B}{\partial\mu_B}$ and $c_{s,\chi}^2=\frac{\partial p_\chi}{\partial\mu_\chi}\bigl/\frac{\partial \varepsilon_\chi}{\partial\mu_\chi}$ in order to express derivatives of the pressures of two components with respect to the corresponding chemical potentials. This expression can be given the form of Eq.~\eqref{IX} with $\eta$ defined as
\begin{eqnarray}
\label{B2}
\eta=\frac{\partial \varepsilon_B}{\partial\mu_B}
\left[\frac{\partial \varepsilon_B}{\partial\mu_B}+
\frac{\partial \varepsilon_\chi}{\partial\mu_\chi}\frac{d\mu_\chi}{d\mu_B}\right]^{-1},
\end{eqnarray}
The thermodynamic identities $n_B=\frac{\partial p_B}{\partial\mu_B}$ and $\varepsilon_B=\mu_B n_B-p_B$ can be used in order to obtain $\frac{\partial \varepsilon_B}{\partial\mu_B}=\mu_B\frac{\partial n_B}{\partial\mu_B}$. We similarly obtain $\frac{\partial \varepsilon_\chi}{\partial\mu_\chi}=\mu_\chi\frac{\partial n_\chi}{\partial\mu_\chi}$. Within stellar interiors chemical potential of two components scale proportionally to each other \citep{PhysRevD.102.063028}. This allows us to conclude that $\frac{d\mu_\chi}{d\mu_B}=\frac{\mu_\chi}{\mu_B}=\xi$. Thus, Eq.~\eqref{B2} becomes
\begin{eqnarray}
\label{B3}
\eta=\frac{\partial n_B}{\partial\mu_B}
\left[\frac{\partial n_B}{\partial\mu_B}+
\xi^2\frac{\partial n_\chi}{\partial\mu_\chi}\right]^{-1}.
\end{eqnarray}
Mechanical stability of BM and DM requires $\frac{\partial n_B}{\partial\mu_B}>0$ and $\frac{\partial n_\chi}{\partial\mu_\chi}$, respectively. In this case $\eta$ by construction is limited to the interval $\eta\in[0,1]$. The lower edge of this interval $\eta=0$ is obtained at $\mu_B=0$, which corresponds to the absence of BM. On the other hand, the case of pure BM is obtained at $\mu_\chi=0$ yielding $\eta=1$.

\newpage
\section{Scan Over Boson Mass and the Interaction Scale}
\label{appC:Scan}
  
   \begin{figure*}[ht!]
    \centering
    \setkeys{Gin}{width=1.05\linewidth}
    \begin{tabularx}{\linewidth}{XXX}
\includegraphics{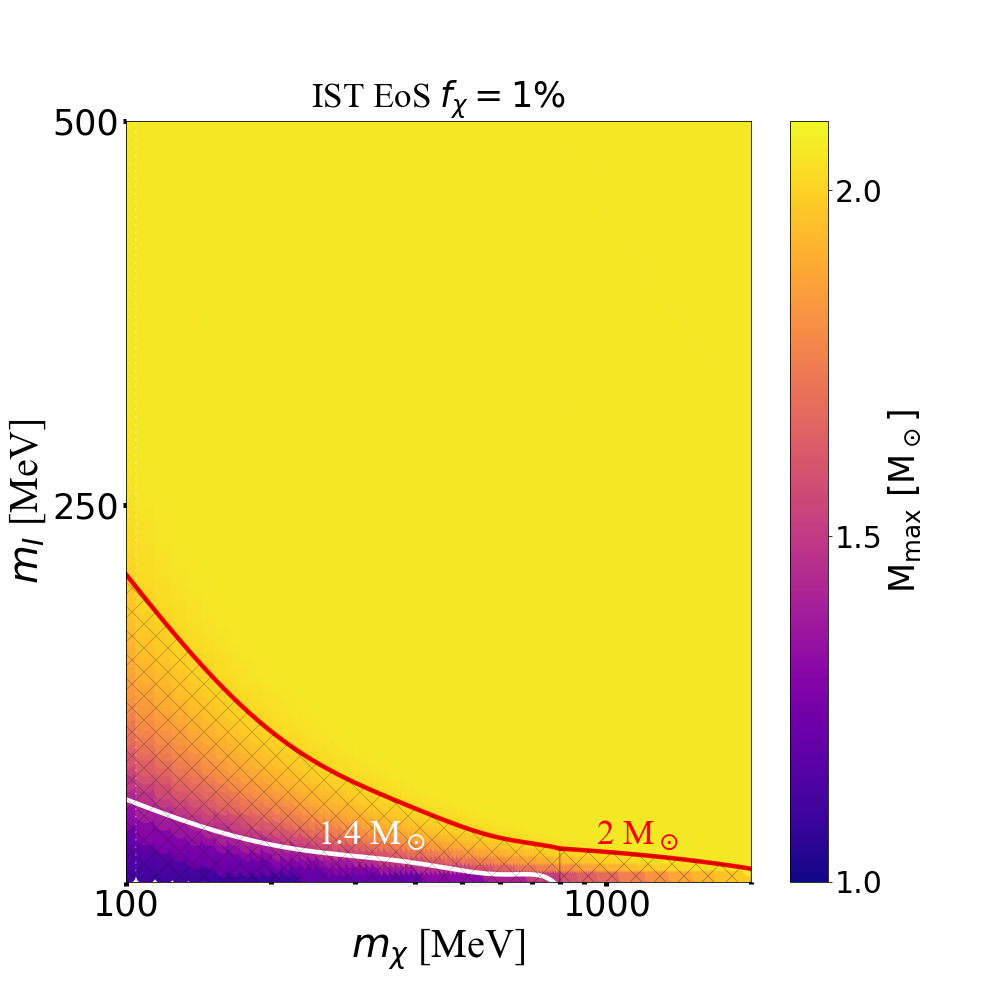}
    &
\hspace{0.15cm} \includegraphics{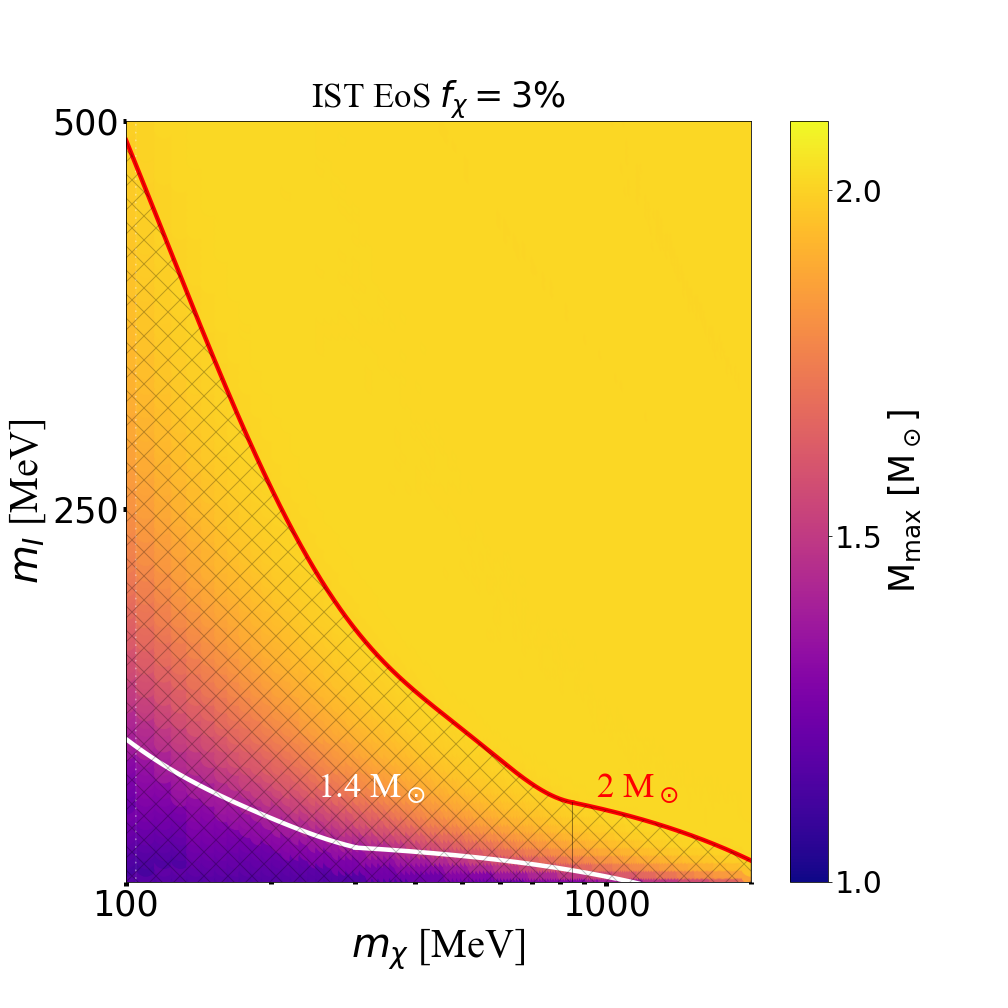}
    &
\hspace{0.15cm} \includegraphics{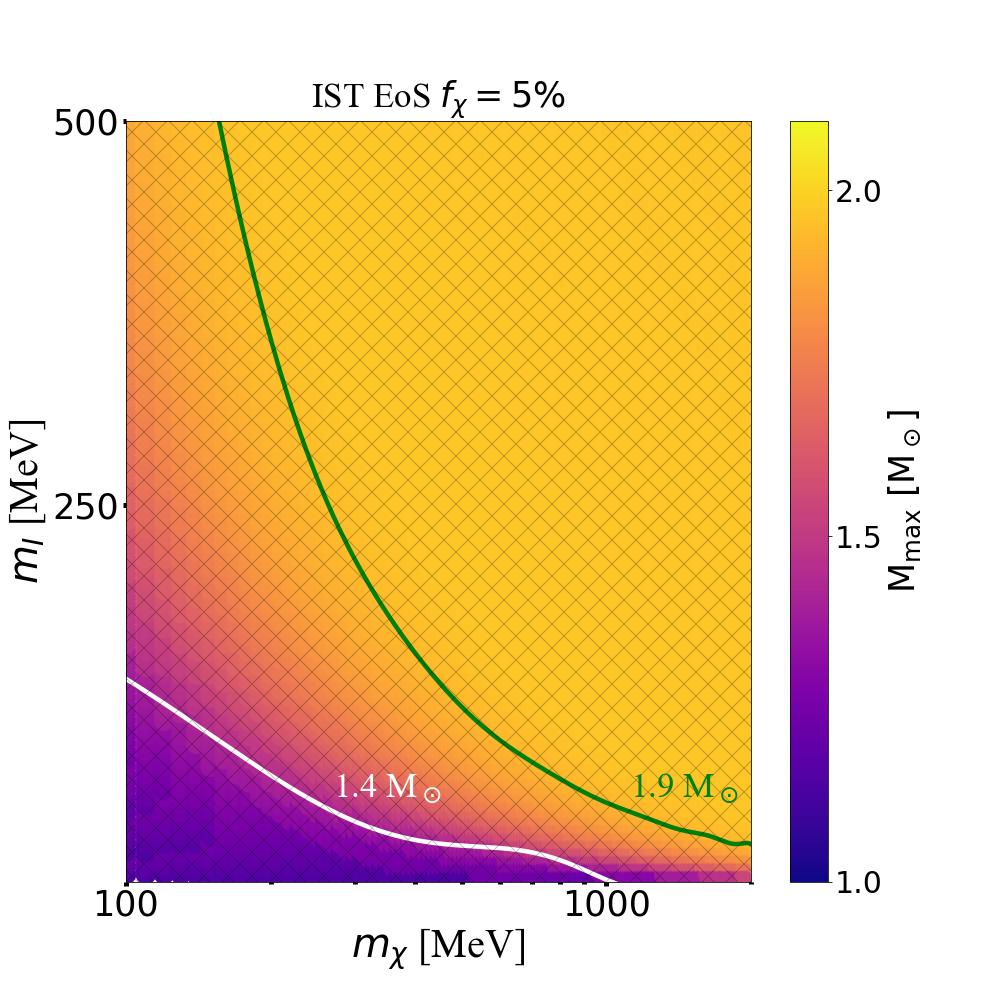}
    \end{tabularx}
        \begin{tabularx}{\linewidth}{XXX}
\includegraphics{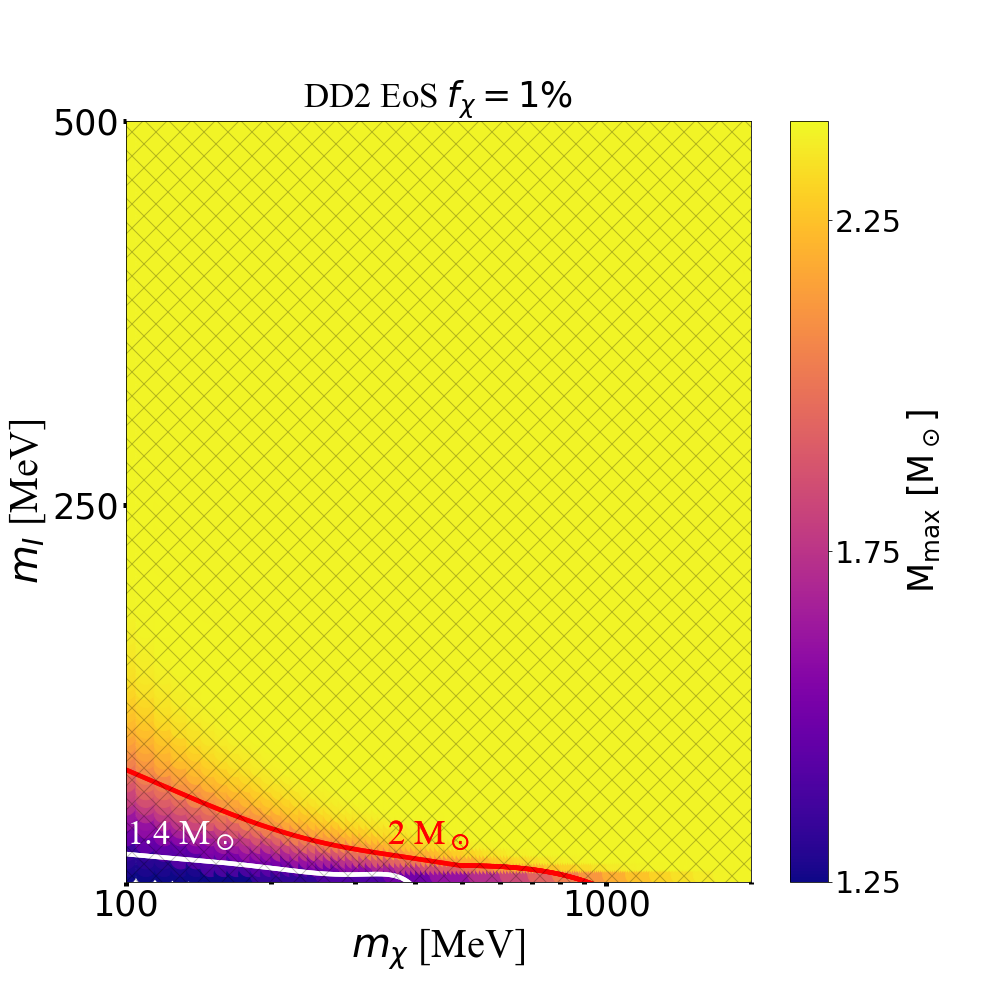}
    &
\hspace{0.15cm} \includegraphics{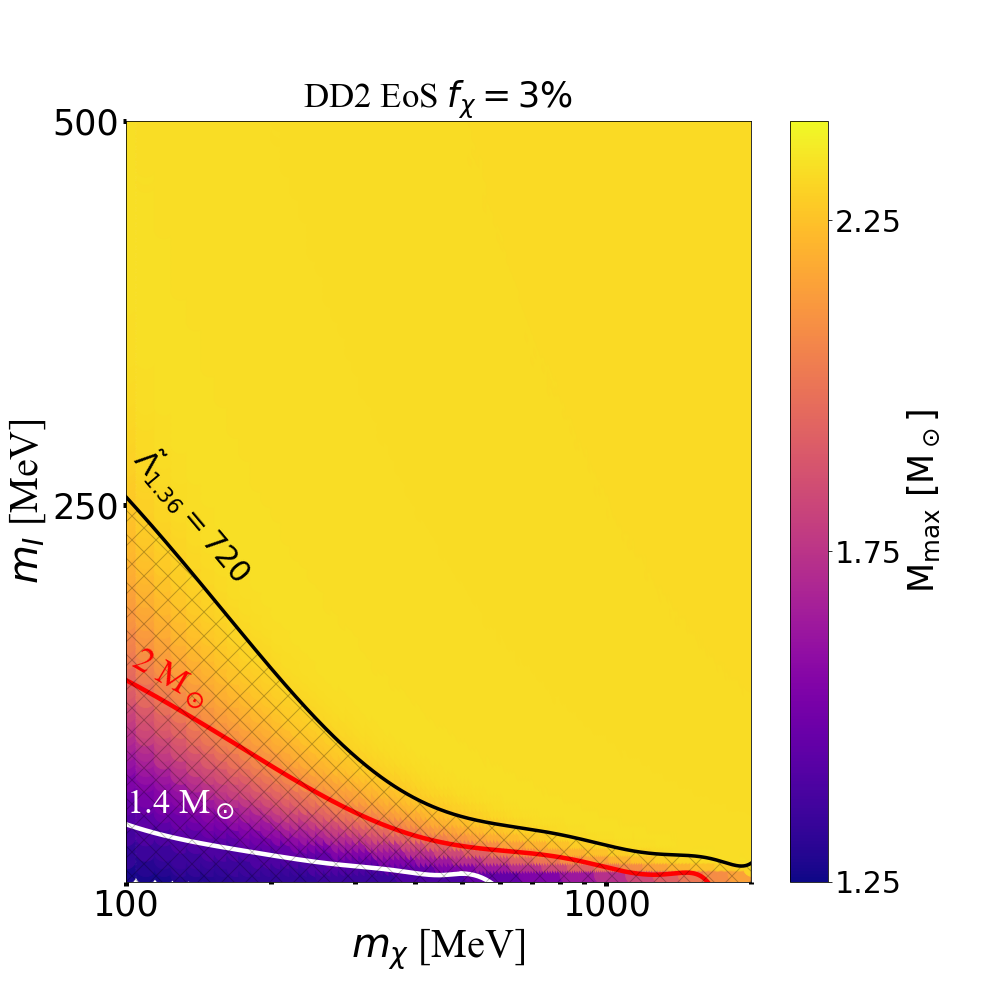}
    &
\hspace{0.15cm} \includegraphics{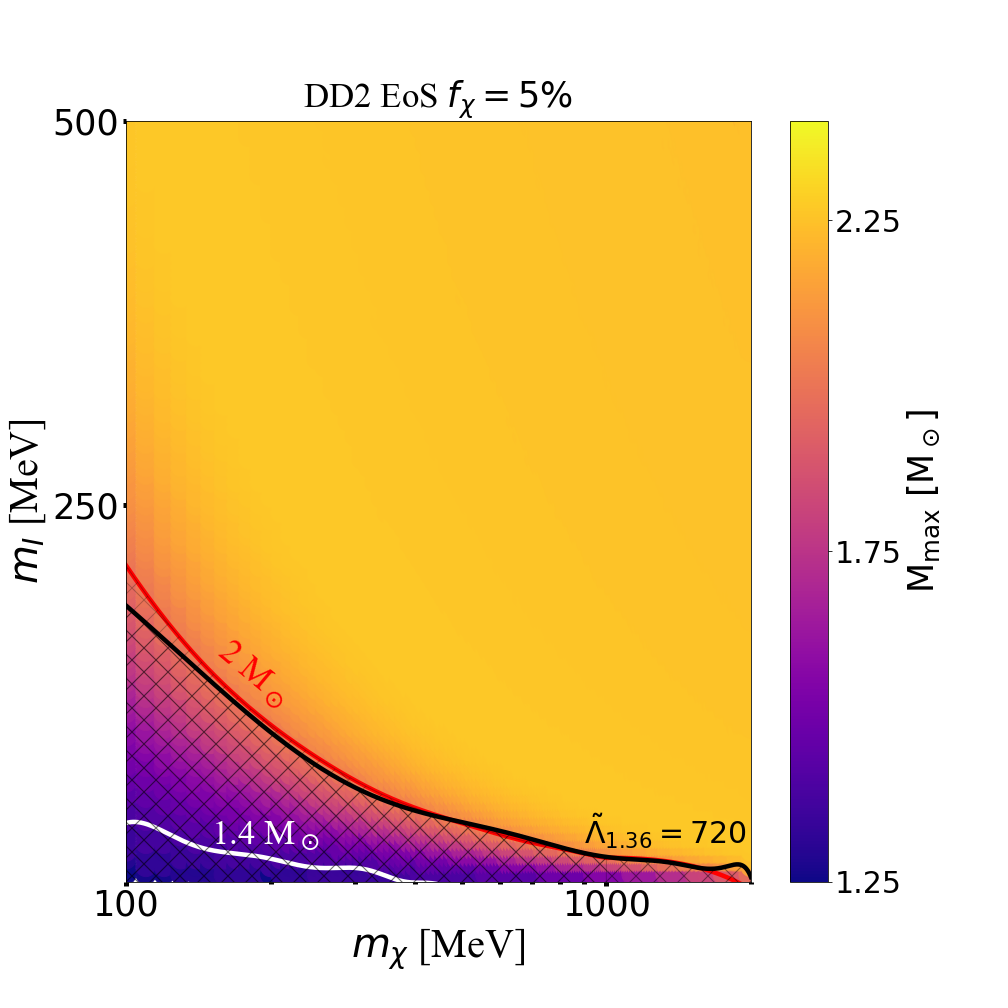}
    \end{tabularx}
            \begin{tabularx}{\linewidth}{XXX}
\includegraphics{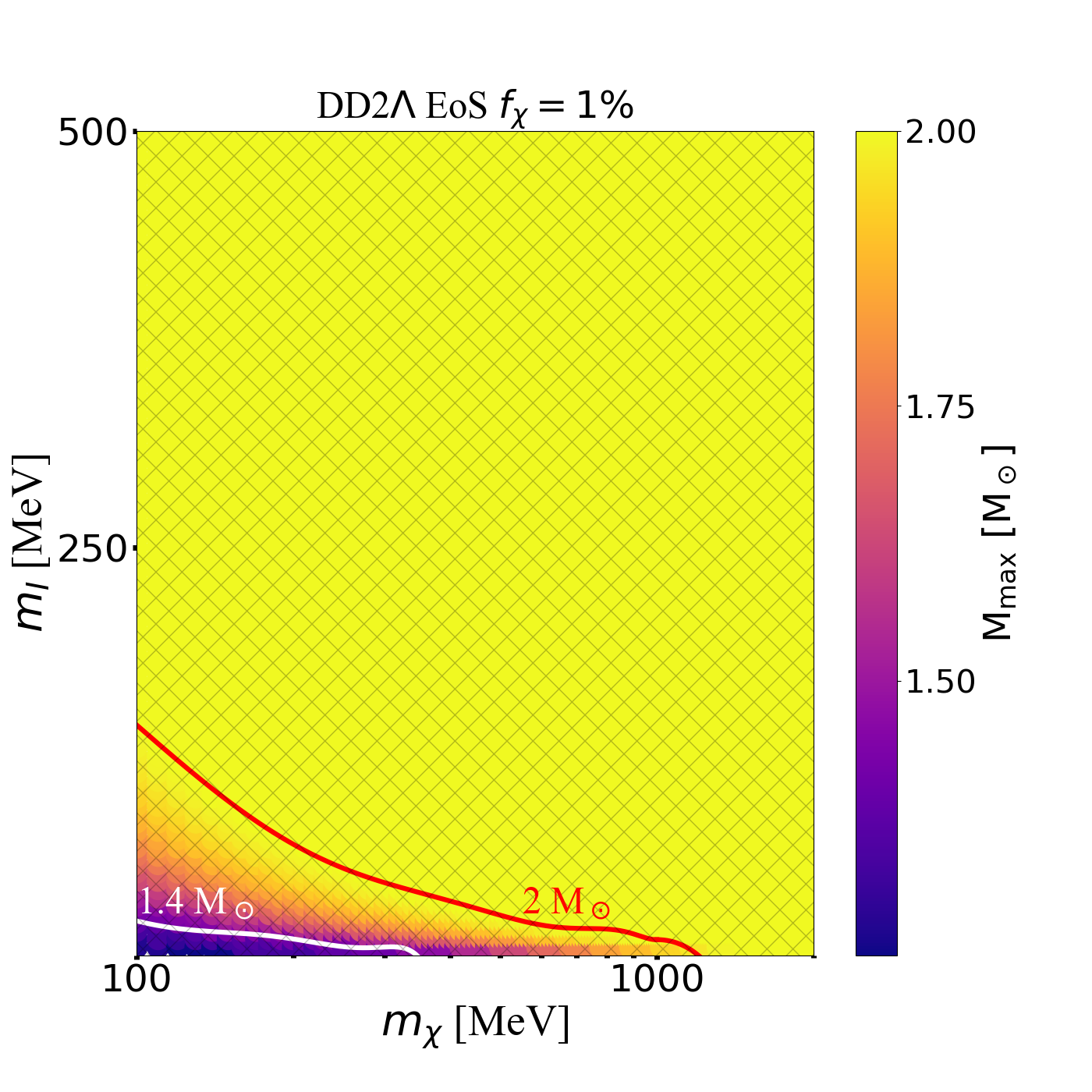}
    &
\hspace{0.15cm} \includegraphics{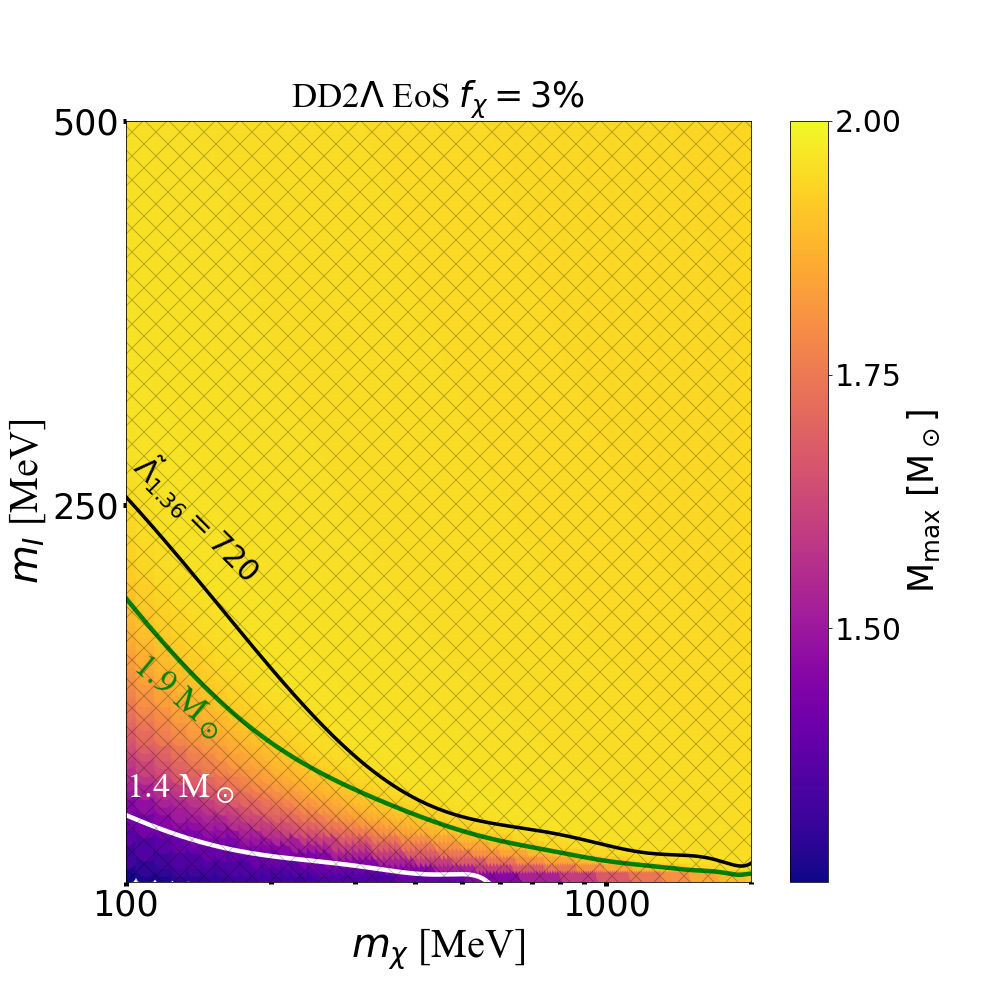}
    &
\hspace{0.15cm} \includegraphics{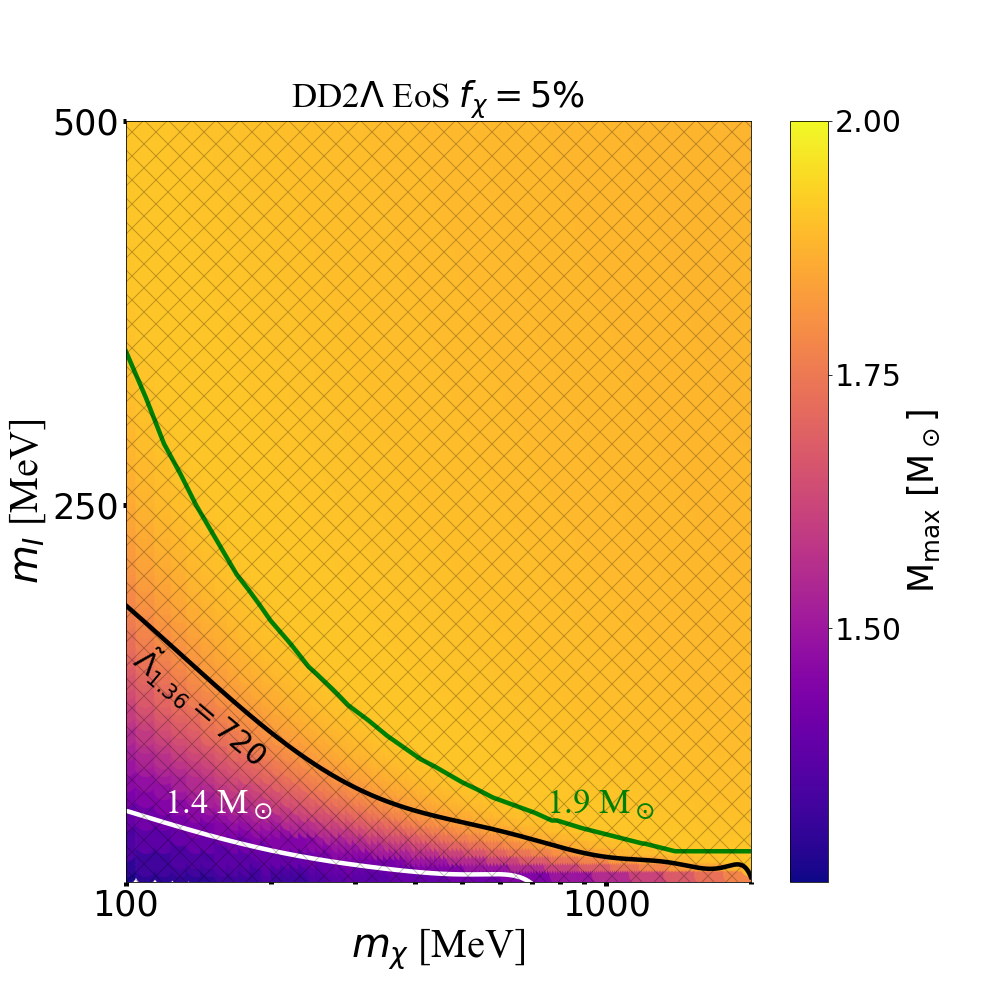}
    \end{tabularx}
    \caption{ {\bf Upper row:}
       Parameter space in the $m_{I}-m_{\chi}$ plane calculated for the IST EoS and different values of DM fraction, 1\% (left panel), 3\% (middle panel), 5\% (right panel). The color represents the total maximum gravitational mass of DM-admixed NSs. The white, and red curves correspond to stars with the total maximum gravitational mass equal to 1.4$~M_{\odot}$, and 2$~M_{\odot}$, respectively. For a DM fraction 5\% (see the right panel) the maximum mass do not reach 2$~M_{\odot}$, thus, we show 1.9$~M_{\odot}$ configurations in green. Shaded areas correspond to the non-allowed regions of parameters, whereas simultaneously the astrophysical and GW constraints are not fulfilled. 
      {\bf Middle row:} The same as on the upper row, but calculated for DD2 EoS. The tidal deformability constraint $\tilde{\Lambda}_{1.36} = 720$  (90\% credible interval) is shown as a black curve. All the area above this curve is consistent with the GW170817 tidal deformability constraint \citep{LIGOScientific:2018hze}. The 1\% DM case is not consistent with the tidal deformability constraint for all values of $m_{I}$ and $m_{\chi}$.
      {\bf Lower row:} The same as on the middle and upper rows, but calculated for DD2$\Lambda$ EoS.
    }
\label{fig:contours}
    \end{figure*}

\bibliography{SIBDM}{}
\bibliographystyle{aasjournal}

\end{document}